\documentclass{aa}
\usepackage{graphicx,natbib}
\bibpunct{(}{)}{;}{a}{}{,}

\newcommand{\lx}{L_\mathrm{3-20}}
\newcommand{\lxint}{L_\mathrm{3-20,int}}
\newcommand{\lh}{L_\mathrm{8-20}}

\newcommand{\ls}{L_\mathrm{2-10}}
\newcommand{\lsint}{L_\mathrm{2-10,int}}

\newcommand{\wx}{W_\mathrm{3-20}}
\newcommand{\nx}{N_\mathrm{3-20}}
\newcommand{\ws}{W_\mathrm{2-10}}

\newcommand{\nh}{N_\mathrm{H}}
\newcommand{\lb}{L_\ast}
\newcommand{\vm}{V_\mathrm{m}}

\newcommand{\lesssim}{\mathrel{\hbox{\rlap{\hbox{\lower4pt\hbox{$\sim$}}}\hbox{$<$}}}}
\newcommand{\gtrsim}{\mathrel{\hbox{\rlap{\hbox{\lower4pt\hbox{$\sim$}}}\hbox{$>$}}}}
\newcommand{\beq}{\begin{equation}}
\newcommand{\eeq}{\end{equation}}
\newcommand{\beqa}{\begin{eqnarray}}
\newcommand{\eeqa}{\end{eqnarray}}

\begin{document}

\title{Statistical properties of local active galactic nuclei
inferred from the RXTE 3--20~keV all-sky survey}

\author{S.Yu. Sazonov\inst{1,2} \and M.G. Revnivtsev\inst{1,2}}

\offprints{sazonov@mpa-garching.mpg.de}

\institute{Max-Planck-Institut f\"ur Astrophysik,
           Karl-Schwarzschild-Str. 1, D-85740 Garching bei M\"unchen,
           Germany
     \and   
           Space Research Institute, Russian Academy of Sciences,
           Profsoyuznaya 84/32, 117997 Moscow, Russia
}
\date{Received 27 January 2004 / Accepted 16 April 2004}

\authorrunning{Sazonov \& Revnivtsev}
\titlerunning{Statistical properties of local AGNs}

\abstract{We compiled a sample of 95 AGNs serendipitously detected in
the 3--20~keV band at Galactic latitude $|b|>10^\circ$ during the RXTE slew
survey (XSS, Revnivtsev et al.), and utilize it to study the
statistical properties of the local population of AGNs, including
the X-ray luminosity function and absorption distribution. We find that 
among low X-ray luminosity ($\lx< 10^{43.5}$~erg~s$^{-1}$)
AGNs, the ratio of absorbed (characterized by intrinsic absorption in
the range $10^{22}$~cm$^{-2}<\nh<10^{24}$~cm$^{-2}$) and unabsorbed
($\nh<10^{22}$~cm$^{-2}$) objects is 2:1, while this ratio drops to
less than 1:5 for higher luminosity AGNs. The
summed X-ray output of AGNs with $\lx>10^{41}$~erg~s$^{-1}$
estimated here is smaller than the earlier estimated total X-ray
volume emissivity in the local Universe, suggesting that a comparable
X-ray flux may be produced together by lower luminosity AGNs,
non-active galaxies and clusters of galaxies. Finally, we present
a sample of 35 AGN candidates, composed of unidentified XSS sources. 
}
\maketitle

\section{Introduction}
\label{intro}

We have recently \citep[][ hereafter Paper 1]{revetal04} taken
advantage of the excellent calibration, moderate field of view (1~deg
radius) and high effective area ($\sim 6000$~sq. cm) of the PCA spectrometer on
board the RXTE observatory to perform  an all-sky survey in
the 3--20~keV band from the data accumulated during satellite
slews in 1996--2002 -- the RXTE slew survey (XSS). For $90$\% of the  
sky at $|b|>10^\circ$, a flux limit for source detection of $2.5\times
10^{-11}$ erg s$^{-1}$ cm$^{-2}$ (3--20~keV) or lower was achieved,
while a combined area of $7\times 10^3$~sq.~deg was sampled to record 
flux levels (for such very large area surveys) below $10^{-11}$ erg
s$^{-1}$ cm$^{-2}$.  

In Paper~1, a catalog comprising 294 X-ray sources detected at
$|b|>10^\circ$ was presented. 236 of these sources were
identified with a single known astronomical object. Of particular
interest are 100 identified active galactic nuclei (AGNs) and 35
unidentified sources. The hard spectra of the latter suggest that many
of them will probably also prove to be AGNs when follow-up observations are
performed. Most of the detected AGNs belong to the local population
($z<0.1$). In addition, the hard X-ray band of the XSS (3--20~keV) as
compared to most previous X-ray surveys, performed at photon energies
below 10~keV, has made possible the detection of a substantial number
of X-ray absorbed AGNs (mostly Seyfert~2 galaxies). These properties
make the XSS sample of AGNs a valuable one for the study of the local
population of AGNs.

In this paper, we carry out a thorough statistical analysis of the
above sample to investigate several key properties of the
local population of AGNs, in particular their distribution in intrinsic
absorption column density ($\nh$) and X-ray luminosity
function. Knowledge of these characteristics provides important
constraints for AGN unification models and synthesis of the cosmic X-ray
background, and is further needed to understand the details of the
accretion-driven growth of supermassive black holes in the nuclei of
galaxies. 

In the course of the paper, we compare our results with previously
published ones. These include the X-ray luminosity function of local
AGNs derived from the HEAO-1/A2 all-sky survey \citep{picetal82}, the
$\nh$ distribution of optically selected Seyfert~2 galaxies
\citep{risetal99} and the evolving with redshift properties of AGNs
inferred largely from medium-sensitivity and deep X-ray surveys
\citep{lafetal02,uedetal03,steetal03}. Finally, we 
assess the contribution of AGNs with luminosities above $\sim
10^{41}$~erg~s$^{-1}$ to the total X-ray volume emissivity in the local
Universe, as estimated by \citet{miyetal94}. 

\section{The sample}
\label{sample}

There are 100 identified AGNs in the XSS catalog, of which 95 make up
the input sample (Table~\ref{agn_table}) for the current study. One source (the
radio galaxy 4C +21.55) was excluded because its redshift is
unknown. Another 4 objects (Mrk 335, Mrk 348, Ton S180 and NGC 1068)
were excluded because they would not have satisfied the 4$\sigma$
detection criterion (in the 3--20~keV energy band) had there  
been no RXTE slews associated with pointed observations of these sources,
i.e. they were not detectable serendipitously during the
survey. Note that NGC 1068 is a Compton thick Seyfert 2 galaxy (i.e. 
having an X-ray spectrum characterized by an intrinsic absorption 
column density $\nh>1.5\times 10^{24}$~cm$^{-2}$), and after its 
removal from the list we are left with just one source of this type 
-- NGC 4945. Given this fact, we restrict the present analysis to AGNs
with $\nh<10^{24}$~cm$^{-2}$, i.e. to Compton thin sources.

\begin{table*}
\caption{AGN sample}
\smallskip

\begin{tabular}{lllrrrrrrrrrrl}
\hline
\hline
\multicolumn{1}{c}{XSS object} &
\multicolumn{1}{c}{Common name} &
\multicolumn{1}{c}{Class$^{\rm c}$} & 
\multicolumn{1}{c}{3--8 keV$^{\rm d}$} & 
\multicolumn{1}{c}{8--20 keV$^{\rm d}$} & 
\multicolumn{1}{c}{$z$} &
\multicolumn{1}{c}{$D^{\rm e}$} & 
\multicolumn{2}{c}{$\log L_{3-20}$$^{\rm f}$} &
\multicolumn{1}{c}{$N_{\rm H}$$^{\rm g}$} & 
\multicolumn{1}{c}{Ref$^{\rm h}$} \\
   
\multicolumn{1}{c}{(J2000.0)} &
\multicolumn{1}{c}{} &     
\multicolumn{1}{c}{} & 
\multicolumn{1}{c}{cnt s$^{-1}$} &       
\multicolumn{1}{c}{cnt s$^{-1}$} &  
\multicolumn{1}{c}{} &
\multicolumn{1}{c}{Mpc} & 
\multicolumn{2}{c}{erg s$^{-1}$} &     
\multicolumn{1}{c}{10$^{22}$ cm$^{-2}$} &
\multicolumn{1}{c}{} \\
\hline
00368+4557   & CGCG 535-012$^{\rm a}$      &    S1 & $0.26\pm0.06$ &
$0.19\pm0.07$ & 0.0480 &      & 43.48 & 43.48 & $12(<26)$     & 1 \\    
01232$-$3502 & NGC 526A          &    S2 & $1.61\pm0.15$ & $1.01\pm0.17$ & 0.0190 &      & 43.42 & 43.44 &   $1.2\pm0.1$ & 3 \\                                  
01236$-$5854 & Fairall 9         &    S1 & $1.26\pm0.10$ & $0.68\pm0.15$ & 0.0460 &      & 44.08 & 44.08 &   $<1$        & 3  \\                                  
02151$-$0033 & MRK 590           &    S1 & $1.95\pm0.39$ &
$1.07\pm0.49$ & 0.0270 &      & 43.79 & 43.79 &   $<1$        & 3 \\                                  
02284+1849   & TEX 0222+185$^{\rm a}$      &   BL? & $1.17\pm0.26$ &
$0.80\pm0.35$ & 2.7000 &      & 48.18 & 48.18 &  $<10^3$       & 1 \\
02360$-$5234 & ESO 198-G024      &    S1 & $1.22\pm0.19$ & $0.25\pm0.24$ & 0.0450 &      & 43.94 & 43.94 &   $<1$        & 3 \\                                  
04182+0056   & 1H 0414+009       &    BL & $0.89\pm0.17$ & $0.22\pm0.19$ & 0.2900 &      & 45.56 & 45.56 &   $<1$        & 3 \\                                  
04273$-$5749 & 1H 0419$-$577     &    S1 & $0.81\pm0.16$ & $0.39\pm0.19$ & 0.1000 &      & 44.57 & 44.58 &   $<1$        & 3 \\                                  
04331+0520   & 3C 120            &  BLRG & $2.94\pm0.20$ & $1.74\pm0.20$ & 0.0330 &      & 44.16 & 44.16 &   $<1$        & 3 \\                                  
05103+1640   & IRAS 05078+1626   &    S1 & $2.08\pm0.20$ & $1.18\pm0.23$ & 0.0180 &      & 43.47 & 43.47 &   $5(<11)$     & 1 \\  
05162$-$0008 & AKN 120           &    S1 & $2.14\pm0.18$ & $1.10\pm0.18$ & 0.0330 &      & 44.00 & 44.00 &   $<1$        & 3 \\                                  
05220$-$4603 & Pic A             &  BLRG & $0.88\pm0.09$ & $0.53\pm0.11$ & 0.0350 &      & 43.69 & 43.69 &   $<1$        & 3 \\                                  
05316+1320   & PKS 0528+134      &    BL & $0.60\pm0.13$ & $0.26\pm0.16$ & 2.1000 &      & 47.53 & 47.53 &   $<1$        & 3 \\                                  
05510$-$3226 & PKS 0548$-$32     &    BL & $1.49\pm0.20$ & $0.62\pm0.25$ & 0.0690 &      & 44.48 & 44.48 &   $<1$        & 3 \\                                  
05518$-$0733 & NGC 2110          &    S2 & $2.29\pm0.18$ & $1.66\pm0.20$ & 0.0076 &      & 42.82 & 42.89 &   $7.0\pm0.6$ & 2 \\                                 
05552+4617   & MCG +8-11-11      &    S1 & $2.37\pm0.27$ &
$0.88\pm0.31$ & 0.0200 &      & 43.56 & 43.56 &   $<1$        & 3 \\                                  
06014$-$5044 & PKS 0558$-$504    &  NLS1 & $0.75\pm0.15$ &
$0.22\pm0.18$ & 0.1400 &      & 44.80 & 44.80 &   $<1$        & 3 \\                                  
06171+7102   & MRK 3             &    S2 & $0.37\pm0.16$ & $0.68\pm0.20$ & 0.0140 &      & 42.99 & 43.40 &  $73\pm3$     & 2 \\                                 
07434+4945   & MRK 79            &    S1 & $1.32\pm0.25$ & $0.76\pm0.36$ & 0.0220 &      & 43.45 & 43.45 &   $<1$        & 4 \\                  
08117+7600   & PG 0804+761       &    S1 & $0.65\pm0.11$ & $0.30\pm0.15$ & 0.1000 &      & 44.47 & 44.47 &   $<1$        & 3 \\                                  
08418+7052   & S5 0836+71        &    BL & $1.64\pm0.12$ & $1.27\pm0.15$ & 2.2000 &      & 48.12 & 48.12 &   $<1$        & 3 \\                                  
09204+1608   & MRK 704           &    S1 & $1.07\pm0.22$ & $0.80\pm0.28$ & 0.0290 &      & 43.65 & 43.65 &   $<1$        & 3 \\                                  
09261+5204   & MRK 110           &  NLS1 & $1.18\pm0.16$ & $0.54\pm0.22$ & 0.0360 &      & 43.80 & 43.81 &   $<1$        & 3 \\                                  
09476$-$3100 & MCG -5-23-16      &    S2 & $6.47\pm0.34$ & $3.91\pm0.31$ & 0.0083 &      & 43.30 & 43.32 &   $2.3\pm0.5$ & 2 \\                                 
10231+1950   & NGC 3227          &    S1 & $2.11\pm0.18$ & $1.38\pm0.18$ & 0.0038 & 20.6 & 42.40 & 42.40 &   $<1$        & 3 \\                                  
11045+3808   & MRK 421           &    BL & $8.99\pm0.12$ & $3.03\pm0.12$ & 0.0310 &      & 44.52 & 44.52 &   $<1$        & 3 \\                                  
11047$-$2341 & 4U 1057$-$21      &    BL & $1.49\pm0.34$ & $0.53\pm0.35$ & 0.1900 &      & 45.41 & 45.41 &   $<1$        & 5 \\                     
11067+7234   & NGC 3516          &    S1 & $3.21\pm0.09$ &
$1.91\pm0.09$ & 0.0088 &      & 43.04 & 43.04 &   $<1$        & 3 \\                                  
11349+6944   & MRK 180           &    BL & $0.42\pm0.06$ &-$0.05\pm0.07$ & 0.0450 &      & 43.34 & 43.34 &   $<1$        & 2 \\                                 
11393$-$3745 & NGC 3783          &    S1 & $4.91\pm0.12$ & $2.93\pm0.18$ & 0.0097 &      & 43.31 & 43.31 &   $<1$        & 3 \\                                  
11417+5910   & SBS 1136+594      &    S1 & $0.42\pm0.07$ & $0.20\pm0.08$ & 0.0600 &      & 43.82 & 43.82 &   $<7$        & 1 \\            
11570+5514   & NGC 3998          & LLAGN & $0.48\pm0.08$ & $0.07\pm0.09$ & 0.0035 & 21.6 & 41.64 & 41.64 &   $<1$        & 3 \\                                  
12032+4424   & NGC 4051          &  NLS1 & $1.49\pm0.07$ & $0.73\pm0.10$ & 0.0023 & 17.0 & 42.04 & 42.04 &   $<1$        & 3 \\                                  
12106+3927   & NGC 4151          &    S1 & $9.47\pm0.10$ & $6.83\pm0.12$ & 0.0033 & 20.3 & 43.08 & 43.17 &   $8.2\pm0.3$ & 2 \\                                 
12164+1427   & PG 1211+143       &  NLS1 & $0.60\pm0.09$ & $0.23\pm0.11$ & 0.0810 &      & 44.22 & 44.22 &   $<1$        & 3 \\                                  
12190+4715   & NGC 4258          &    S2 & $0.81\pm0.07$ & $0.51\pm0.10$ & 0.0015 &  6.8 & 41.04 & 41.11 &   $7.0\pm1.0$ & 6 \\                  
12206+7509   & MRK 205           &    S1 & $0.37\pm0.06$ & $0.15\pm0.08$ & 0.0700 &      & 43.88 & 43.88 &   $<1$        & 3 \\                                  
12260+1248   & NGC 4388          &    S2 & $4.34\pm0.10$ & $2.62\pm0.10$ & 0.0084 & 16.8 & 42.65 & 42.93 & $40\pm10$     & 7 \\                   
12288+0200   & 3C 273            &    BL & $6.78\pm0.18$ & $4.29\pm0.12$ & 0.1600 &      & 45.98 & 45.98 &   $<1$        & 3 \\                                  
12351$-$3948 & NGC 4507          &    S2 & $0.29\pm0.11$ & $0.57\pm0.13$ & 0.0120 &      & 42.72 & 43.06 &  $58.7\pm1.5$ & 2 \\                                 
12408$-$0516 & NGC 4593          &    S1 & $1.05\pm0.17$ & $0.62\pm0.19$ & 0.0090 &      & 42.57 & 42.57 &   $<1$        & 3 \\                                  
12553$-$2633 & CTS M12.22$^{\rm a}$        &    S1 & $0.46\pm0.09$ & $0.21\pm0.11$ & 0.0580 &      & 43.82 & 43.82 &   $<13$       & 1 \\  
12565$-$0537 & 3C 279            &    BL & $0.77\pm0.11$ & $0.48\pm0.16$ & 0.5400 &      & 46.25 & 46.25 &   $<1$        & 3 \\                                  
13073$-$4926 & NGC 4945          &    S2 & $0.31\pm0.10$ & $0.31\pm0.11$ & 0.0019 &  5.2 & 40.92 & 41.66 & $220$         & 8 \\                       
13085$-$4018 & ESO 323-G077$^{\rm b}$   &    S1 & $0.31\pm0.12$ & $0.46\pm0.14$ & 0.0150 &      & 42.86 & 43.21 & $55\pm33$     & 1 \\     
13253$-$4302 & Cen A             &  NLRG &$13.00\pm0.17$ & $9.52\pm0.18$ & 0.0018 &  4.9 & 42.00 & 42.11 &  $11.0\pm1.0$ & 2 \\                                 
13312$-$2502 & ESO 509-G038      &    S1 & $0.26\pm0.06$ & $0.16\pm0.07$ & 0.0260 &      & 42.90 & 42.90 &  $7(<20)$     & 1 \\               
13354$-$3414 & MCG -6-30-15      &    S1 & $3.08\pm0.07$ & $1.65\pm0.07$ & 0.0077 &      & 42.89 & 42.89 &   $<1$        & 9 \\
13420$-$1432 & NPM1G $-$14.0512  &  NLS1 & $0.47\pm0.08$ & $0.26\pm0.10$ & 0.0420 &      & 43.57 & 43.57 &   $<1$        & 1 \\     
13492$-$3020 & IC 4329A          &    S1 & $7.30\pm0.08$ & $4.21\pm0.08$ & 0.0160 &      & 43.91 & 43.91 &   $<1$        & 9 \\
13530+6916   & MRK 279           &    S1 & $2.14\pm0.07$ & $1.04\pm0.08$ & 0.0310 &      & 43.94 & 43.94 &   $<1$        & 3 \\                                  
13578$-$4214 & PKS 1355$-$41     &   RLQ & $0.39\pm0.08$ & $0.29\pm0.10$ & 0.3100 &      & 45.42 & 45.42 & $31(<57)$     & 1 \\         
14132$-$0311 & NGC 5506          &    S2 & $6.90\pm0.14$ & $3.72\pm0.14$ & 0.0062 &      & 43.06 & 43.09 &   $2.6\pm0.2$ & 2 \\                                 
14176$-$4910 & PKS 1416$-$49$^{\rm a}$     &    RG & $0.53\pm0.10$ & $0.28\pm0.13$ & 0.0920 &      & 44.33 & 44.33 &   $4(<17)$    & 1 \\     
14181+2514   & NGC 5548          &    S1 & $3.58\pm0.14$ & $2.04\pm0.14$ & 0.0170 &      & 43.65 & 43.66 &   $<1$        & 3 \\                                  
14194$-$2606 & ESO 511-G030      &    S1 & $1.11\pm0.09$ & $0.71\pm0.09$ & 0.0220 &      & 43.39 & 43.39 &   $<1$        & 3 \\                                  
14278+4240   & H 1426+428        &    BL & $1.17\pm0.09$ & $0.55\pm0.12$ & 0.1300 &      & 44.98 & 44.98 &   $<1$        & 3 \\                                  
15042+1046   & MRK 841           &    S1 & $0.95\pm0.14$ & $0.32\pm0.17$ & 0.0360 &      & 43.67 & 43.67 &   $<1$        & 3 \\                                                  
\end{tabular}

\label{agn_table}
\end{table*}

\setcounter{table}{0}
\begin{table*}
\caption{--continued}
\smallskip

\begin{tabular}{lllrrrrrrrrrl}
\hline
\hline
\multicolumn{1}{c}{XSS object} &
\multicolumn{1}{c}{Common name} &
\multicolumn{1}{c}{Class$^{\rm c}$} & 
\multicolumn{1}{c}{3--8 keV$^{\rm d}$} & 
\multicolumn{1}{c}{8--20 keV$^{\rm d}$} & 
\multicolumn{1}{c}{$z$} &
\multicolumn{1}{c}{$D^{\rm e}$} & 
\multicolumn{2}{c}{$\log L_{3-20}$$^{\rm f}$} &
\multicolumn{1}{c}{$N_{\rm H}$$^{\rm g}$} & 
\multicolumn{1}{c}{Ref$^{h}$} \\
   
\multicolumn{1}{c}{(J2000.0)} &
\multicolumn{1}{c}{} &     
\multicolumn{1}{c}{} & 
\multicolumn{1}{c}{cnt s$^{-1}$} &       
\multicolumn{1}{c}{cnt s$^{-1}$} &  
\multicolumn{1}{c}{} &
\multicolumn{1}{c}{Mpc} & 
\multicolumn{2}{c}{erg s$^{-1}$} &     
\multicolumn{1}{c}{10$^{22}$ cm$^{-2}$} &
\multicolumn{1}{c}{} \\          
\hline
15128$-$0858 & PKS 1510$-$08     &    BL & $0.59\pm0.15$ & $0.51\pm0.20$ & 0.3600 &      & 45.78 & 45.78 &   $<1$        & 3 \\                                   
15348+5750   & MRK 290           &    S1 & $0.68\pm0.08$ & $0.37\pm0.09$ & 0.0300 &      & 43.43 & 43.43 &   $<1$        & 3 \\                                   
15478$-$1350 & NGC 5995          &    S2 & $1.06\pm0.15$ & $0.40\pm0.17$ & 0.0250 &      & 43.41 & 43.41 &   $<1$        & 11 \\                   
16253+5201   & SBS 1624+514$^{\rm a}$      &  BLRG & $0.25\pm0.07$ & $0.20\pm0.08$ & 0.1800 &      & 44.70 & 44.70 &   $<1$        & 1 & \\
16536+3951   & MRK 501           &    BL & $9.47\pm0.15$ & $5.20\pm0.15$ & 0.0330 &      & 44.66 & 44.66 &   $<1$        & 3 \\                                   
17169$-$6232 & NGC 6300          &    S2 & $0.45\pm0.11$ & $0.52\pm0.15$ & 0.0037 & 14.3 & 41.62 & 41.83 &  $25.2\pm1.0$ & 2 \\                                  
17272+5025   & I Zw 187          &    BL & $0.60\pm0.12$ & $0.35\pm0.15$ & 0.0550 &      & 43.93 & 43.93 &   $<1$        & 2 \\                                     
17276$-$1359 & PDS 456           &   RQQ & $0.59\pm0.11$ & $0.17\pm0.15$ & 0.1800 &      & 44.93 & 44.93 &   $<1$        & 12 \\                     
17413+1851   & 4C +18.51         &   RLQ & $1.56\pm0.19$ & $0.66\pm0.25$ & 0.1900 &      & 45.45 & 45.45 &  $<4.6$       & 1 \\                
18196+6454   & H 1821+643        &   RQQ & $0.91\pm0.13$ & $0.22\pm0.16$ & 0.3000 &      & 45.60 & 45.61 &   $<1$        & 3 \\                                   
18348+3238   & 3C 382            &  BLRG & $2.29\pm0.14$ & $1.40\pm0.16$ & 0.0590 &      & 44.58 & 44.58 &   $<1$        & 2 \\                                  
18362$-$5918 & Fairall 49        &    S2 & $1.00\pm0.06$ & $0.36\pm0.07$ & 0.0200 &      & 43.20 & 43.24 &   $4.2\pm0.5$ & 2 \\                                  
18376$-$6511 & ESO 103-G035      &    S2 & $1.28\pm0.07$ & $0.99\pm0.09$ & 0.0130 &      & 43.12 & 43.33 &  $27.1\pm0.6$ & 2 \\                                  
18408+7947   & 3C 390.3          &  BLRG & $1.67\pm0.10$ & $0.97\pm0.10$ & 0.0560 &      & 44.39 & 44.39 &   $<1$        & 3 \\                                   
18449$-$6204 & ESO 140-G043      &    S1 & $0.78\pm0.12$ & $0.35\pm0.14$ & 0.0140 &      & 42.79 & 42.79 &   $<1$        & 13 \\                  
18494$-$7829 & H 1846$-$786      &    S1 & $1.15\pm0.10$ & $0.58\pm0.12$ & 0.0740 &      & 44.46 & 44.46 &   $<1$        & 3 \\                                   
19202$-$5849 & ESO 141$-$G055    &    S1 & $1.38\pm0.13$ & $0.80\pm0.16$ & 0.0360 &      & 43.91 & 43.91 &   $<1$        & 3 \\                                   
19592+6505   & 1ES 1959+650      &    BL & $4.11\pm0.11$ & $1.11\pm0.13$ & 0.0470 &      & 44.52 & 44.53 &   $<1$        & 2 \\                                  
20085$-$4833 & PKS 2005$-$489    &    BL & $3.81\pm0.18$ & $2.66\pm0.18$ & 0.0710 &      & 44.99 & 44.99 &   $<1$        & 2 \\                                  
20404+7521   & 4C +74.26         &   RLQ & $1.09\pm0.13$ & $0.64\pm0.17$ & 0.1000 &      & 44.73 & 44.73 &   $<1$        & 3 \\                                   
20441$-$1042 & MRK 509           &    S1 & $3.13\pm0.28$ & $1.68\pm0.28$ & 0.0340 &      & 44.20 & 44.20 &   $<1$        & 2 \\                                  
20501$-$5646 & IC 5063           &    S2 & $0.71\pm0.13$ & $0.66\pm0.16$ & 0.0110 &      & 42.77 & 43.01 &  $31.5\pm1.6$ & 2 \\                                  
21128+8216   & S5 2116+81        &  BLRG & $0.47\pm0.12$ & $0.35\pm0.16$ & 0.0840 &      & 44.25 & 44.25 &   $<1$        & 1 \\      
21310$-$6219 & IRAS F21325$-$6237&    S1 & $0.51\pm0.10$ & $0.18\pm0.12$ & 0.0590 &      & 43.85 & 43.85 &  $<13$        & 1 \\              
21583$-$3004 & PKS 2155$-$304    &    BL & $5.04\pm0.23$ & $2.07\pm0.14$ & 0.1200 &      & 45.52 & 45.52 &   $<1$        & 3 \\                                   
22013$-$3147 & NGC 7172          &    S2 & $1.18\pm0.17$ & $0.80\pm0.20$ & 0.0086 &      & 42.65 & 42.76 &  $12.6\pm0.7$ & 2 \\                                  
22023+4226   & BL Lac            &    BL & $0.62\pm0.09$ & $0.37\pm0.12$ & 0.0690 &      & 44.15 & 44.15 &   $<1$        & 3 \\                                   
22082$-$4708 & NGC 7213          &    S1 & $1.13\pm0.16$ & $0.57\pm0.19$ & 0.0060 &      & 42.22 & 42.22 &   $<1$        & 3 \\                                   
22355$-$2601 & NGC 7314          &    S2 & $2.16\pm0.20$ & $1.21\pm0.25$ & 0.0047 & 18.3 & 42.28 & 42.28 &   $<1$        & 3 \\                                   
22363$-$1230 & MRK 915           &    S1 & $0.79\pm0.17$ & $0.51\pm0.21$ & 0.0240 &      & 43.32 & 43.32 &  $9(<22)$     & 1 \\              
22423+2958   & AKN 564           &  NLS1 & $1.13\pm0.14$ & $0.35\pm0.14$ & 0.0250 &      & 43.41 & 43.42 &   $<1$        & 3 \\                                   
22539$-$1735 & MR 2251$-$178     &   RQQ & $2.10\pm0.20$ & $1.37\pm0.20$ & 0.0640 &      & 44.63 & 44.63 &   $<1$        & 12 \\                     
23033+0858   & NGC 7469          &    S1 & $1.70\pm0.20$ & $0.91\pm0.10$ & 0.0160 &      & 43.27 & 43.27 &   $<1$        & 3 \\                                   
23040$-$0834 & MRK 926           &    S1 & $1.61\pm0.17$ & $1.06\pm0.21$ & 0.0470 &      & 44.23 & 44.23 &   $<1$        & 3 \\                                   
23073+0447   & PG 2304+042       &    S1 & $0.55\pm0.10$ & $0.36\pm0.12$ & 0.0420 &      & 43.67 & 43.67 & $12(<26)$     & 1 \\               
23178$-$4236 & NGC 7582          &    S2 & $0.58\pm0.16$ & $0.41\pm0.16$ & 0.0530 &      & 43.95 & 44.07 & $14.8\pm1.2$  & 2 \\                                  
15103$-$2042 & IRAS 15091$-$2107 &  NLS1 & $0.61\pm0.10$ & $0.07\pm0.12$ & 0.0450 &      & 43.60 & 43.60 &   $<1$        &10 \\                                     
\hline
\end{tabular}

$^{\rm a}$ Previous X-ray detection only below 2~keV in the ROSAT
All-Sky Survey.  
 
$^{\rm b}$ Previous X-ray detection only marginally by HEAO-1/A2
\citep{deletal90}.
 
$^{\rm c}$ RQQ -- radio-quite quasar, RLQ -- radio-loud quasar, BL
-- blasar (BL Lac object or flat-spectrum radio quasar), S1 --  
Seyfert 1 galaxy (types 1, 1.2 and 1.5), NLS1 -- narrow-line Seyfert 1
galaxy, S2 -- Seyfert 2 galaxy (types 1.8, 1.9 and 2), RG -- radio
galaxy, BLRG -- broad-line radio galaxy, NLRG --  narrow-line radio
galaxy, LLAGN -- low luminosity AGN. 

$^{\rm d}$ Errors are 1$\sigma$ statistical uncertainties.

$^{\rm e}$ If no value is given, then the distance is calculated from
the redshift

$^{\rm f}$ Left and right columns give the observed and intrinsic
luminosities in the 3--20~keV band, respectively.

$^{\rm g}$ Errors are 1$\sigma$ statistical uncertainties, upper
limits are 1$\sigma$ if based on the XSS hardness ratio and more
conservative in other cases.

$^{\rm h}$ Quoted $\nh$ value is estimated or adopted from: (1) XSS
catalog (Paper~1); (2) RXTE/PCA pointing observations; (3)
TARTARUS/ASCA database; (4) \citealt{turpou89}; (5)
\citealt{woletal98}; (6) \citealt{youwil04}; (7) \citealt{risaliti02};
(8) \citealt{matetal00}; (9) \citealt{reynolds97}; (10)
\citealt{awaetal91}; (11) ASCA public data; (12) \citealt{reetur00};
(13) \citealt{ghosou92}.
    
\end{table*}

For each object in the sample a detailed AGN class is adopted 
from the XSS catalog, which in turn mostly follows the classification
of the NED database. The sample includes 18 blazars. Their emission,
including the X-rays, is collimated in our direction, which makes this
class distinctly different from normal, emission-line AGNs. We have decided to
consider the poorly studied source TEX 0222+185 a blazar due to its
extraordinary inferred X-ray luminosity ($\sim
10^{48}$~erg~s$^{-1}$). The remaining 77 sources are non-blazar AGNs of
various types, mostly Seyfert galaxies. Note that no strict division
is drawn here between Seyfert galaxies and quasars; typically, AGNs
designated as Seyferts (or radio galaxies) and quasars have an X-ray
luminosity below and above $10^{44.5}$~erg~s$^{-1}$, respectively. 

In the non-blazar subsample, 60 objects are optically classified as
type 1 AGNs and 7 of these are narrow-line Seyfert 1 galaxies
(NLS1). We can directly infer from these numbers that 
in the local Universe NLS1 galaxies make up $\sim 10$\% of hard
X-ray (3--20~keV) selected type 1 AGNs. This result fits well in the
picture summarized by \cite{grupe00} that NLS1 galaxies appear
significantly enhanced (reaching $\sim 40$\%) in soft X-ray selected
samples compared with hard X-ray selected ones.  Radio loud AGNs (radio
galaxies and radio-loud quasars) amount to $10/60\sim 15$\% of type 1
objects in our sample, which is consistent with the well-known fraction ($\sim
10$\%) of radio loud objects among optically selected quasars
\citep[e.g.][]{iveetal02}. We note that it would be wrong to estimate
here in the same straightforward way the proportion of unabsorbed and absorbed
AGNs, which is one of the primary goals of this study, since our sample
is biased against the latter type due to the sampled space volume
decreasing with increasing $\nh$ for a given intrinsic luminosity. The
corresponding accurate calculation will be done in \S\ref{nh_dist}.

The information given in Table~\ref{agn_table} for each AGN includes
the measured count rates in two energy bands 3--8~keV and 8--20~keV
together with their 1$\sigma$ statistical uncertainties. We point out
that these count rates have been obtained by averaging over multiple
slews performed at random times during the period 1996--2002. It is
important to note that in contrast to the original XSS catalog, the
uncertainties quoted here do not take into account RXTE slews related
to pointings at the sources. The current sample is thus effectively
serendipitous. As was noted above, the corresponding correction has
led to the removal of 4 AGNs from the sample. All of the 
presented AGNs are detected at a more than 4$\sigma$ confidence level in
the 3--20~keV band. It should be noted that at the faintest fluxes,
source confusion may affect the count rate estimation. In Paper~1 a
relevant threshold was estimated as $4\sigma_{\rm
conf}=0.5$~cnt~s$^{-1}$ for the 3--20~keV band. Since only 3 of the 
AGNs in our sample have measured count rates below this limit, the
overall effect of confusion on the sample is definitely negligible.

Next, two types of luminosity in the observer's 3--20~keV
band\footnote{Defining the luminosities in the rest-frame
3--20~keV band makes essentially no difference, since the estimated
$k$-correction $|\Delta\log L|<0.1$ for all the sources in the sample.} 
 are given for each source. The observed luminosity, $\lx$, is
calculated from the measured 3--20~keV count rate (the sum of the
3--8~keV and 8--20~keV count rates) by taking into account the
spectral response of the RXTE/PCA instrument and assuming a power-law
spectrum of photon index $\Gamma=1.8$ with a low-energy cutoff due to
intrinsic absorption (see \S\ref{nh} below). The intrinsic luminosity
$\lxint$ is then found by correcting the observed luminosity for the
intrinsic absorption. We also note that the quoted luminosities for
the only AGN in our sample with $\nh>10^{24}$~cm$^{-2}$ -- NGC~4945 --
should be regarded as crude estimates because the X-ray spectrum of
this Compton thick source is poorly described \citep{guaetal00} by the
photoabsorbed power law model adopted here.

The luminosity distances were computed from the known redshifts
assuming a cosmology with $(H_0, \Omega_{\rm m}, \Omega_\Lambda)=
(75$~km~s$^{-1}$~Mpc$^{-1}$, 0.3, 0.7). For 10 nearby 
sources ($z\lesssim 0.01$), the distances from the 
Nearby Galaxies Catalogue \citep{tully88} were adopted.
 
\begin{figure}
\centering
\includegraphics[width=\columnwidth]{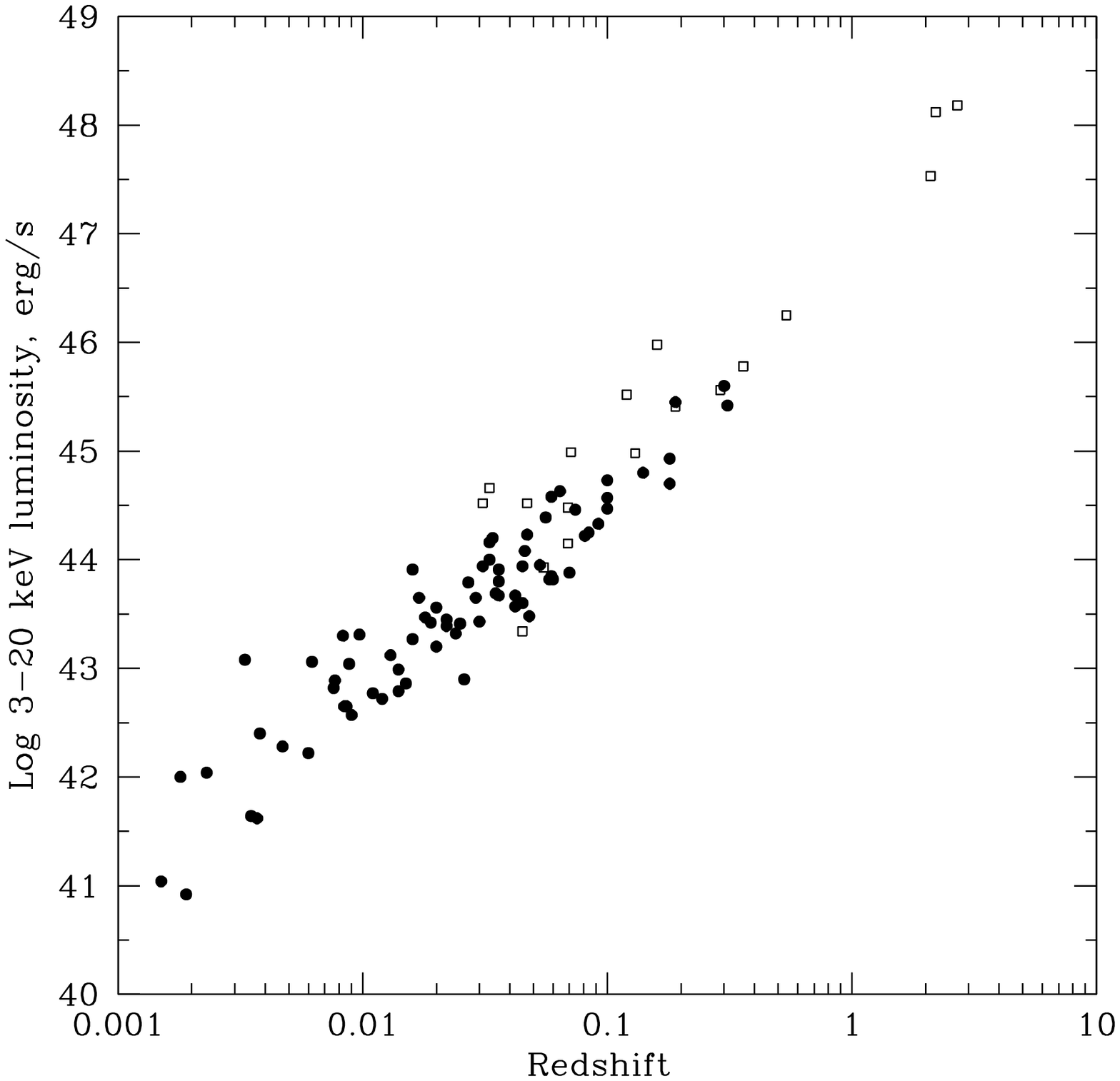}
\caption{Distribution in redshift and luminosity of 95 known AGNs
detected in the RXTE slew survey, including 77 
emission-line AGNs (solid circles) and 18 blazars (open squares).  
\label{z_lum}
} 
\end{figure}    
We show in Fig.~\ref{z_lum} the location of our AGNs on the
redshift--luminosity plane. One can see that the distribution in
observed luminosity is extremely broad, spanning 7 (5) orders of
magnitude if the blazars are included or not. We are
effectively probing the local Universe ($z<0.1$). These properties
combined with the hard X-ray range (3--20~keV) distinguish our survey
from others.

\subsection{Absorption column density}
\label{nh}

\begin{figure}
\centering
\includegraphics[width=\columnwidth]{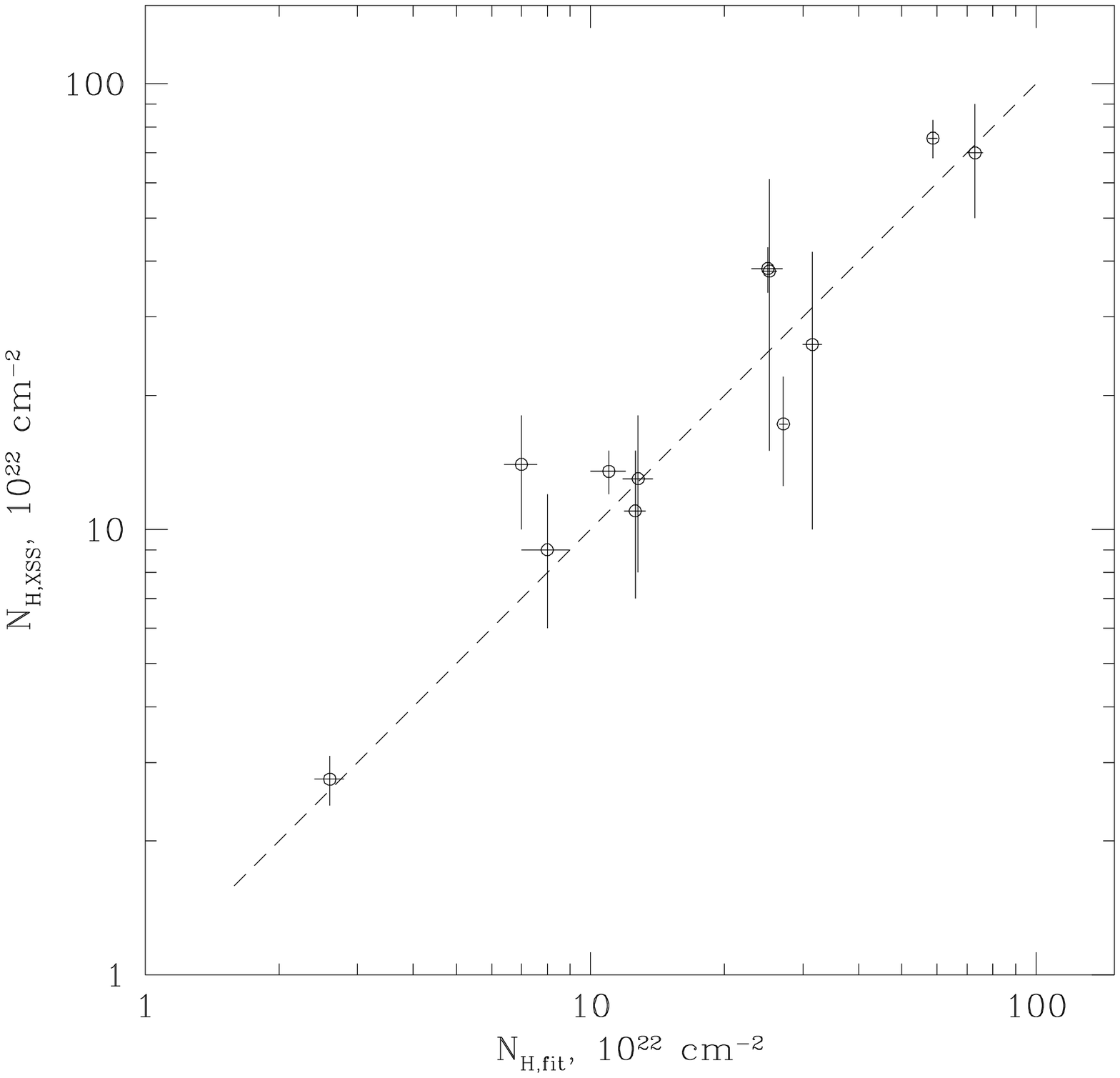}
\caption{Comparison of the absorption columns estimated from
the XSS 8--20~keV/3--8~keV count rate ratio, assuming a
$\Gamma=1.8$ power-law intrinsic spectrum, with those 
derived through spectral analaysis of RXTE/PCA pointed
observations for Seyfert~2 galaxies with a significant XSS detection
of neutral absorption. The error bars are 1$\sigma$ statistical uncertainties. 
\label{nh_nh}
} 
\end{figure}

Another important property of AGNs is intrinsic absorption
column density ($\nh$). It plays crucial roles in the AGN
unification paradigm and in the study of the cosmic X-ray
background. It should be noted that throughout this paper we ignore the
contribution of Compton scattering to intrinsic extinction and
consider photoabsorption only. This is justified because we do not
consider Compton thick sources and since the effect of Compton
scattering on the spectrum is expected to be small (typically less than
$25$\% of the flux density at photon energies from 3--20~keV) for
$\nh< 10^{24}$~cm$^{-2}$ \citep{yaqoob97,matetal99} and further tends
to be counteracted by the presence of a reflected spectral component
\citep{matetal00,risaliti02} neglected here.

For all of our identified or candidate (see \S\ref{unident}) AGNs, the
absorption column can be estimated to a first approximation from the
ratio of the measured count rates in the 8--20~keV and 3--8~keV bands,
assuming a $\Gamma=1.8$ power-law intrinsic spectrum and taking
into account the source redshift (if known). The above spectral slope is
typical for Seyfert galaxies and quasars, as is known from previous
studies \citep[e.g.][]{reynolds97} and also follows from our own
analysis of pointed RXTE observations. Since NLS1 galaxies typically
have somewhat softer ($\Gamma\approx 2.2$) and unabsorbed spectra
\citep[e.g.][]{leighly99}, the above procedure is expected to give the correct
result ($\nh=0$) also for objects of this type. In this work, we do not
distinguish column densities below $10^{22}$~cm$^{-2}$, and therefore
possible source-to-source variations of the order of 0.2 in the
intrinsic power-law index should have no effect on our
results. Interstellar absorption is similarly unimportant, since the
XSS sources are located at $|b|>10^\circ$.

For most of the identified AGNs, we have been able to improve the above
crude estimate of the absorption column either by analysing
the spectral data of relevant pointed RXTE observations, or adopting
$\nh$ values from the TARTARUS/ASCA database or the literature.  
Data of RXTE/PCA pointed observations were reduced using the FTOOLS/LHEASOFT
5.3 package. The spectral modelling was done with XSPEC. AGN spectra
were fitted by a simplistic model consisting of a power-law
component photoabsorbed by a column of neutral material and a
fluorescent neutral iron emission line at energy 6.4 keV, if the
latter was required by the fit [model $zphabs*(power+gaus)$ in
XSPEC]. For almost all the absorbed AGNs, the spectral fitting yielded
values of the photon index $\Gamma\approx 1.8$, except in the
well-known case of NGC 4151 ($\Gamma\approx 1.4$) (see
\citealt{schwar02} and references therein). The obtained 
$\nh$ values were then compared with those estimated from the XSS catalog. 
The result of this comparison (Fig.\ref{nh_nh}) demonstrates that the
photoabsorption column density can be robustly estimated from the XSS
hardness ratio. We note that systematic uncertainties of the RXTE/PCA
energy response do not allow one to measure absorption columns with an accuracy
better than $(0.5-1)\times 10^{22}$~cm$^{-2}$.

\begin{figure}
\centering
\includegraphics[width=\columnwidth]{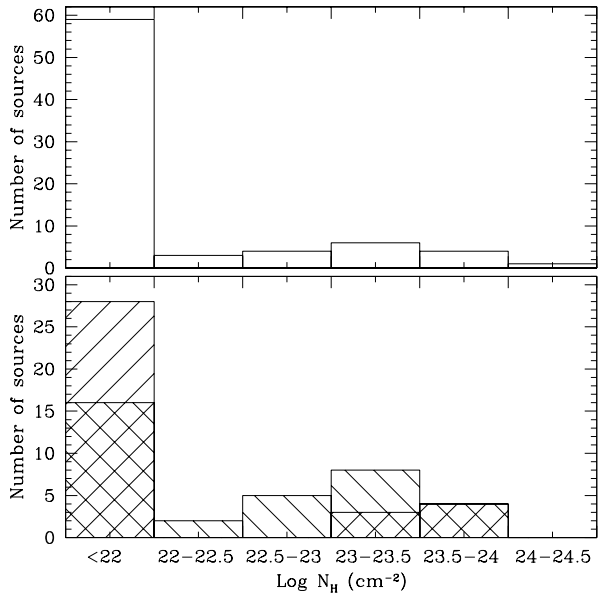}
\caption{{\bf Upper panel:} Observed distribution in intrinsic
absorption column density ($\nh$) of emission-line AGNs in our
sample. {\bf Lower panel:} Estimated $\nh$ distribution of the unidentified
XSS sources, AGN candidates. Two different histograms are presented, one
computed by adopting $\nh=0$ in those cases where the measured $\nh$
value is smaller than the 1$\sigma$ statistical uncertainty, and the
other computed using the best-fit $\nh$ values, regardless of
their errors.
\label{nhdist_obs}
} 
\end{figure}
 
In Fig.~\ref{nhdist_obs} we show the observed distribution of our
non-blazar AGNs in $\nh$. In obtaining this plot, we adopted $\nh=0$
in the 11 cases where only an upper limit exceeding $10^{22}$~cm$^{-2}$
is available. This is a reasonable assumption because all these
sources are optically classified as type 1 AGNs. One can see that most
of the sources in the sample have $\nh<10^{22}$~cm$^{-2}$. However,
the remaining 18 AGNs provide us with enough information to reconstruct  
the physical $\nh$ distribution of AGNs (see \S\ref{nh_dist}). As
expected, almost all of our AGNs with $\nh>10^{22}$~cm$^{-2}$ are
optically classified as Seyfert 2 galaxies. One exception is NGC~4151,
a well-known Seyfert 1 galaxy. We here point out another possible one -- ESO
323-G077. Should our crude estimate (based on the hardness ratio) 
$\nh=(6\pm 3)\times 10^{23}$~cm$^{-2}$ be confirmed by future  X-ray
spectroscopic observations, this source will present a unique example
of a Seyfert 1 galaxy with a heavily absorbed X-ray spectrum. We note that our
identification of this source in Paper~1 was based on its optical brightness
($m_V=13.2$~mag) and the suggestion by \citet{schetal03}, motivated by
spectropolarimetry of the object, that ESO 323-G077 may be 
a transition case between Seyfert 1 and Seyfert 2 galaxies due to its
orientation.  

\subsection{Unidentified sources}
\label{unident}

There are 35 sources in the XSS catalog that remain unidentified
(Table~\ref{noid_table}). We argued in Paper~1 that a large,
probably dominant fraction of these sources are previously 
unknown AGNs. The main argument was that the distribution of their
effective 3--20~keV spectral slopes is similar to that of the
identified AGNs in the catalog. Being of great interest and waiting for
identification, the unidentified sources present the largest source of
incompleteness for the present statistical study. We note that a
negligibly small number (4) of the unidentified sources have measured
count rates falling below the $4\sigma_{\rm conf}$ confusion limit
described above.

\begin{table*}
\caption{Sample of unidentified sources}
\smallskip

\begin{tabular}{lrrrl}
\hline
\hline
\multicolumn{1}{c}{XSS object} &
\multicolumn{1}{c}{3--8 keV$^{\rm a}$} & 
\multicolumn{1}{c}{8--20 keV$^{\rm a}$} & 
\multicolumn{1}{c}{$N_{\rm H}$$^{\rm b}$} & 
\multicolumn{1}{c}{Possible X-ray associations}\\

\multicolumn{1}{c}{(J2000.0)} &
\multicolumn{1}{c}{cnt s$^{-1}$} &       
\multicolumn{1}{c}{cnt s$^{-1}$} &  
\multicolumn{1}{c}{10$^{22}$ cm$^{-2}$} &
\multicolumn{1}{c}{} \\
\hline
00050$-$6904 & $0.48\pm0.04$ & $0.22\pm0.05$ & $<1.6$      & \\
00564+4548   & $0.71\pm0.04$ & $0.37\pm0.05$ & $1.6(<4.4)$ & 1RXS J005528.0+461143 \\
01023$-$4731 & $0.40\pm0.08$ & $0.33\pm0.11$ & $22\pm19$   & 1H 0102$-$469 \\
02087$-$7418 & $0.58\pm0.09$ & $0.50\pm0.11$ & $22\pm13$   & \\
05054$-$2348 & $0.69\pm0.19$ & $0.92\pm0.23$ & $48\pm25$   & \\
05188+1823   & $0.61\pm0.15$ & $0.43\pm0.18$ & $14(<30)$   & \\
12270$-$4859 & $1.32\pm0.13$ & $0.68\pm0.16$ & $<4.3$      & 1RXS J122758.8$-$485343 \\
12303$-$4232 & $0.48\pm0.09$ & $0.29\pm0.11$ & $6(<16)$    & 1RXS J123212.3$-$421745 \\
12389$-$1614 & $0.93\pm0.11$ & $0.51\pm0.13$ & $3(<9)$     & \\
13563$-$7342 & $0.54\pm0.10$ & $0.25\pm0.13$ & $<12$       & 1H 1342-733 \\
14101$-$2936 & $0.26\pm0.07$ & $0.31\pm0.09$ & $43\pm26$   & \\
14138$-$4022 & $0.52\pm0.09$ & $0.24\pm0.11$ & $<6$        & \\
14239$-$3800 & $0.77\pm0.12$ & $0.39\pm0.15$ & $<6$        & 1RXS J142149.8$-$380901 \\
14241$-$4803 & $0.33\pm0.05$ & $0.05\pm0.05$ & $<1$        & 1RXS J142148.7$-$480420 \\
14353$-$3557 & $0.34\pm0.08$ & $0.20\pm0.10$ & $6(<18)$    & \\
14408$-$3815 & $0.74\pm0.07$ & $0.28\pm0.09$ & $<1$        & 1RXS J144037.4$-$384658 \\  
14495$-$4005 & $0.42\pm0.05$ & $0.19\pm0.06$ & $<1.8$      & \\   
15076$-$4257 & $0.73\pm0.05$ & $0.24\pm0.06$ & $<1$        & \\
15153$-$3802 & $0.25\pm0.06$ & $0.09\pm0.07$ & $<12$       & \\ 
15179$-$3559 & $0.12\pm0.05$ & $0.13\pm0.06$ & $30(<70)$   & \\
15360$-$4118 & $0.27\pm0.06$ & $0.17\pm0.07$ & $8(<21)$    & \\     
16049$-$7302 & $0.41\pm0.09$ & $0.31\pm0.11$ & $16(<33)$   & \\    
16151$-$0943 & $0.57\pm0.14$ & $0.25\pm0.16$ & $<16$       & 1RXS J161519.2$-$093618, 1H 1613$-$097 \\
16537$-$1905 & $0.83\pm0.15$ & $0.58\pm0.17$ & $13(<27)$   & 1H 1652$-$180 \\ 
17223$-$7301 & $0.38\pm0.08$ & $0.14\pm0.10$ & $<11$       & 1RXS J171850.0$-$732527 \\     
17309$-$0552 & $0.67\pm0.15$ & $0.36\pm0.18$ & $3(<20)$    & 1RXS J173021.5$-$055933, 1H 1726$-$058 \\
17413$-$5354 & $0.43\pm0.09$ & $0.18\pm0.10$ & $<11$       & \\
17459+1115   & $0.27\pm0.23$ & $0.13\pm0.28$ & $<50$       & 1RXS J174527.8+110837 \\ 
17576$-$4534 & $0.49\pm0.10$ & $0.31\pm0.13$ & $7(<19)$    & \\
18236$-$5616 & $0.32\pm0.09$ & $0.51\pm0.12$ & $59\pm27$   & \\
18381$-$5653 & $0.33\pm0.13$ & $0.36\pm0.16$ & $35\pm32$   & \\
18486$-$2649 & $0.74\pm0.14$ & $0.32\pm0.17$ & $<11$       & \\
19303$-$7950 & $0.70\pm0.12$ & $0.34\pm0.16$ & $<7$        & 1RXS J194944.6$-$794519 \\
19459+4508   & $0.38\pm0.10$ & $0.38\pm0.12$ & $30\pm22$   & \\
21354$-$2720 & $0.41\pm0.11$ & $0.32\pm0.13$ & $19(<39)$   & 1RXS J213445.2$-$272551, 1H 2132$-$277 \\
\hline
\end{tabular}

$^{\rm a}$ Errors are 1$\sigma$ statistical uncertainties.
  
$^{\rm b}$ Errors and upper limits are 1$\sigma$.

\label{noid_table}
\end{table*}

Assuming that the unidentified sources are AGNs and that
their intrinsic spectrum is a power law with $\Gamma=1.8$, we
can estimate intrinsic absorption column densities for them from the
measured hardness ratios, exactly as we did before for the known AGNs.
Since the candidate AGNs belong to the faint tail of the XSS catalog,
their estimated aborption columns have large associated statistical
uncertainties $\delta\nh$, in fact in many cases only an upper limit
can be given (see Table~\ref{noid_table}). We have therefore built two 
different $\nh$ distributions, one by adopting $\nh=0$ in
those cases where $\nh<\delta\nh$, and another by using the best-fit
$\nh$ values in all cases. The true distribution likely lies between 
these two approximations. As shown in Fig.~\ref{nhdist_obs}, if the
unidentified sources are AGNs, they are similarly or somewhat more absorbed on 
average than our identified AGNs. We shall return to this point in
\S\ref{nh_dist}. We point out that in the above analysis we assumed that
$z=0$ for the unidentified sources, therefore the inferred 
column densities may be somewhat underestimated.

\subsection{Unresolved sources}
\label{groups}

Our AGN sample may be additionally incomplete due to the presence in
the XSS catalog of 12 sources associated with 2 or 3 unresolved
astronomical objects including at least one AGN. Assuming that all 
sources in each of these groups contribute equally to the measured flux and
applying the 4$\sigma$ detection criterion to the individual sources,
we estimate that about 5 detectable AGNs are probably missing due to
this manifestation of source confusion.

\subsection{Sample completeness}
\label{compl}

It follows from the preceeding discussion that our input sample of 95 AGNs
probably misses up to $\approx 41$ AGNs (including 4C +21.55 with unknown
redshift) meeting the XSS detection criterion, mostly as a result of
the incomplete identification of the catalog. We can make allowance 
for this fact in the subsequent analysis (particularly when
reconstructing the AGN luminosity function in \S\ref{lum_func}) by
introducing a completeness factor of $95/136\approx 70$\%, assuming
that the luminosity distribution of unidentified AGNs is similar to
that of the identified ones. The above value should be considered a lower
limit because some of the 
unidentified sources are probably not AGNs. We may further introduce
similar coefficients (lower limits) for the northern and southern
hemispheres: $45/(45+4+3)\approx 87$\% and $50/(50+31+3)\approx
60$\%. Our AGN sample is thus highly complete in the northern
sky, and we shall take advantage of this fact later in the
paper. Furthemore, our knowledge of the sample completeness is not limited
by the above coefficients, in fact the crude information available on
the X-ray spectra of the unidentified sources is utilized below in the
investigation of the absorption distribution of AGNs.

\subsection{Comparison with the Piccinotti sample}
\label{picci}

Our AGN sample includes 28 of the 35 sources composing the well-known HEAO-1/A2
sample of AGNs \citep{picetal82,maletal99}, based on a complete survey of the
$|b|>20^\circ$ sky down to a limiting flux of $3.1\times 
10^{-11}$~erg~cm$^{-2}$~s$^{-1}$ in the 2--10~keV band. This degree of
overlap of the two catalogs is consistent with our survey being more
sensitive (for 90\% of the sky at $|b|>10^\circ$ an equivalent
2--10~keV sensitivity of $2\times 10^{-11}$~erg~cm$^{-2}$~s$^{-1}$ or
better is achieved) combined with the fact that AGNs are known to be
variable by a factor $\sim 2$ on a time scale of years.

\section{Distribution of AGNs in intrinsic absorption column density}
\label{nh_dist}

\begin{figure}
\centering
\includegraphics[width=\columnwidth]{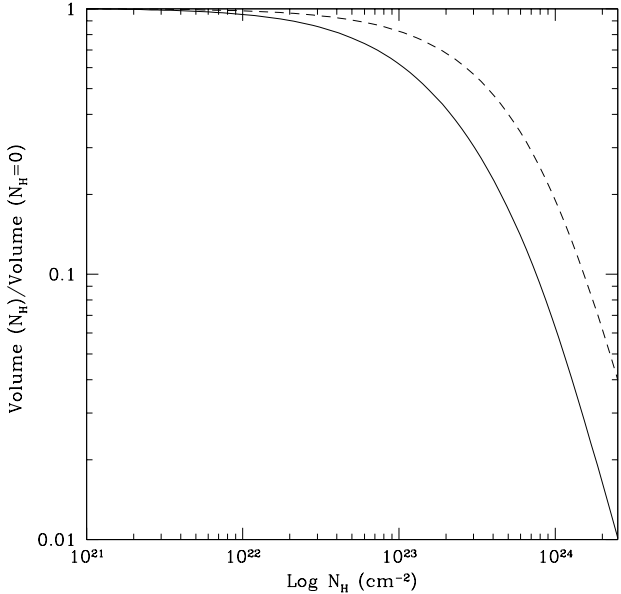}
\caption{Relative volume of space probed by the RXTE slew survey with
respect to AGNs of given intrinsic (unabsorbed) luminosity with a
$\Gamma=1.8$ power-law intrinsic spectrum as a function of absorption
column density. The solid and dashed lines correspond to the 3--20~keV
and 8--20~keV bands, respectively.
\label{vol_nh}
} 
\end{figure}

We now utilize the AGN sample defined in the preceeding section to study
the distribution of local AGNs in intrinsic absorption column density,
$f(\nh)$. To this end, we employ the maximum likelihood technique, as
described below. We note that calculations using the $1/\vm$ method
lead to very similar results, but the maximum likelihood method allows
us to estimate the errors of the output parameters more reliably,
given the relatively small number of absorbed AGNs in our sample.

Assuming that $f(\nh)$ is independent of the intrinsic AGN
luminosity, we model $f(\nh)$ in the range $\nh<10^{24}$~cm$^{-2}$ by a
step-function  
\beq
f(\nh)=f_{i}\,\,\, {\rm if}\,\, 
N_{{\rm H},i}^{\rm min}<\nh<N_{{\rm H},i}^{\rm max}, 
\eeq
where $i=1...5$, $N_{{\rm H},i}^{\rm min}=0$, $10^{22}$, $10^{22.5}$,
$10^{23}$, $10^{23.5}$~cm$^{-2}$ and $N_{{\rm H},i}^{\rm max}=10^{22}$, 
$10^{22.5}$, $10^{23}$, $10^{23.5}$, $10^{24}$~cm$^{-2}$. 
By definition, $\sum_{i=1}^5 f_{i}=1$, and therefore any 4 out of the
5 $f$s may be chosen as the free parameters. 

We find the best-fit model by minimizing the maximum likelihood
estimator, defined as follows:
\beq
L=-2\sum_{j}\ln\frac{f(N_{{\rm H},j})
\vm(L_{\mathrm{3-20,int},j},N_{{\rm H},j})} 
{\int f(\nh)\vm(L_{\mathrm{3-20,int},j},\nh)\,d\nh},
\label{like_nh}
\eeq
where $j$ goes through each AGN in an input sample, and
$\vm(L_{\mathrm{3-20,int},j},\nh)$ is the space volume over which the
$j$th AGN with its intrinsic luminosity $L_\mathrm{3-20,int}$ could be
detected by the survey if its spectrum were a $\Gamma=1.8$ power law
absorbed by gas column $\nh$. The integration in equation~(\ref{like_nh}) is  
carried out from $\nh=0$ to $10^{24}$~cm$^{-2}$.

The relative survey volume, computed from the RXTE/PCA 
energy response and XSS exposure map (see Fig.~1--Fig.~3 in Paper~1),
is shown as a function of $\nh$ in 
Fig.~\ref{vol_nh}. One can see that $\vm$ begins to decrease
noticeably above $\nh=10^{23}$~cm$^{-2}$ and drops 15 times by
$10^{24}$~cm$^{-2}$, as the low-energy cutoff in the X-ray 
spectrum reaches $\sim 10$~keV. This causes the paucity of Compton
thick sources in our catalog. Also shown in Fig.~\ref{vol_nh} is a
similar plot obtained for the hard (8--20~keV) subband, where the
survey sensitivity is less affected by intrinsic absorption. We shall
use the latter dependence in \S\ref{8_20}.

\begin{figure}
\centering
\includegraphics[width=\columnwidth]{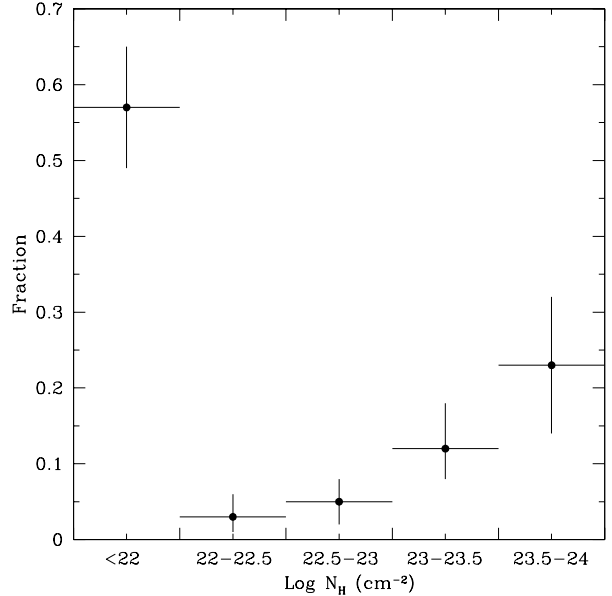}
\caption{Physical distribution in absorption column density of
Compton-thin emission-line AGNs inferred from the XSS AGN sample. The 
error bars in this and subsequent figures represent 1$\sigma$
statistical uncertainties.
\label{nhdist_all}
} 
\end{figure}

The parameter errors are estimated using a standard
procedure: for each of the 4 fitted $f_{i}$, 1$\sigma$ lower
and upper limits are derived that correspond to an increase of $L$ by 1 from
its minimum value while the other parameters are adjusted to
minimize $L$. The confidence region for the 5th (not used in the
fitting) $f_{i}$ is obtained by repeating the procedure, considering
this fraction a free parameter instead of one of the others. 

Performing an analysis along the above lines for our sample of 77
emission-line AGNs we obtain the $\nh$ distribution shown in
Fig.~\ref{nhdist_all}. We find that $57\pm 8$\% of AGNs with
$\nh<10^{24}$~cm$^{-2}$ are unabsorbed ($\nh<10^{22}$~cm$^{-2}$) and
also that more absorbed ($10^{23}$~cm$^{-2}<\nh<10^{24}$~cm$^{-2}$)
sources are $4.1^{+2.9}_{-1.6}$ times as abundant as less absorbed
($10^{22}$~cm$^{-2}<\nh<10^{23}$~cm$^{-2}$) ones.

\begin{figure}
\centering
\includegraphics[width=\columnwidth]{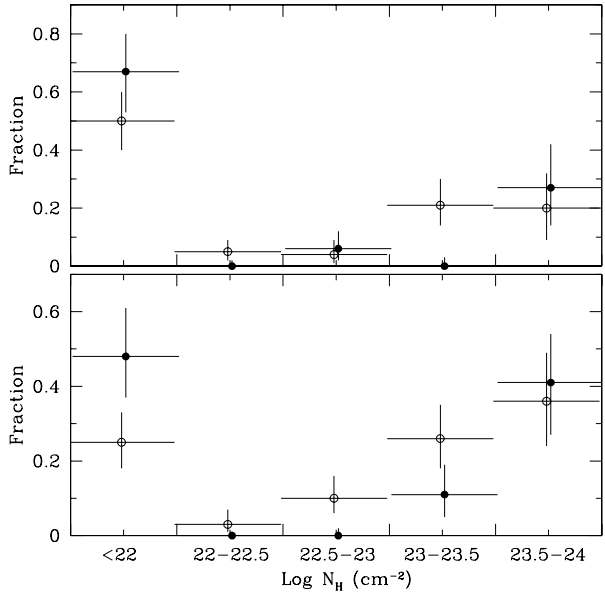}
\caption{{\bf Upper panel:} AGN absorption column density distribution
inferred from the northern (filled circles) and southern (open
circles) subsamples of identified AGNs. {\bf Lower panel:} $\nh$  distribution
estimated for the unidentified XSS sources, assuming that they are
AGNs. One estimate (filled circles) is obtained by adopting $\nh=0$ in
those cases where the measured $\nh$ value is smaller than the 1$\sigma$
statistical uncertainty, another one (open circles) is based on the measured
$\nh$ values. The different sets of data points are slightly shifted
relative to each other long the horizontal axis for better visibility.
\label{nhdist_ns_noid}
} 
\end{figure}

It is important to assess the effect of incompleteness of our AGN sample on
the above result. As a first test, we compare in
Fig.~\ref{nhdist_ns_noid} the $\nh$ distributions inferred from the
northern and southern subsamples. These distributions are apparently
similar and consistent with that obtained for the whole sample
(Fig.~\ref{nhdist_all}), and we recall that our AGN sample is highly
complete in the northern hemisphere. As a further test, we can build a
$\nh$ distribution for our unidentified sources by assuming that all
of them are AGNs with an intrinsic $\Gamma=1.8$ power-law spectrum. By adopting
that $\nh=0$ if $\nh<\delta\nh$ and alternatively using the $\nh$
values given in Table~\ref{noid_table} in all cases, we obtain two
distributions, shown in Fig.~\ref{nhdist_ns_noid}, that likely bound the 
true one. This distribution is also not significantly different
from that derived for the known AGNs. We conclude that the effect of
incompleteness on the distribution shown in Fig.~\ref{nhdist_all} is
small.

\subsection{Luminosity dependence}
\label{nh_dist_lum}

\begin{figure}
\centering
\includegraphics[width=\columnwidth]{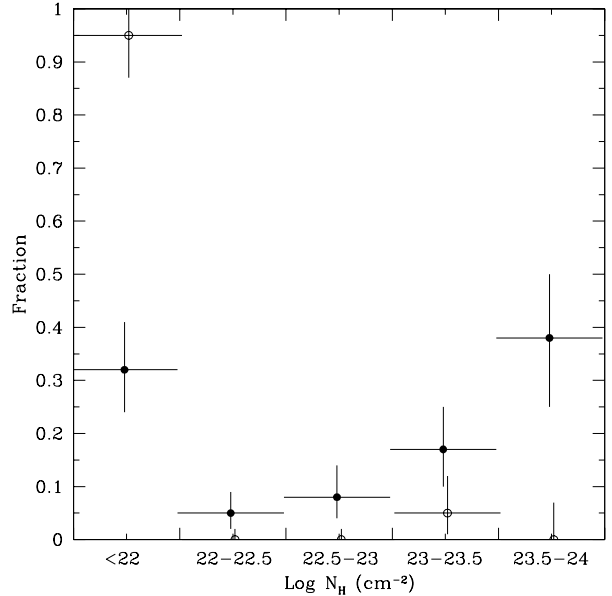}
\caption{Absorption column density distribution of emission-line AGNs
with luminosity $\lx<10^{43.5}$~erg~s$^{-1}$ (filled circles) and
$\lx>10^{43.5}$~erg~s$^{-1}$ (open circles).
\label{nhdist_lum}
} 
\end{figure}

We now wish to investigate whether the $\nh$ distribution depends on
luminosity. To this end, we repeat the above analysis separately 
for non-blazar AGNs with intrinsic luminosity $\lxint$ below and above
$10^{43.5}$~erg~s$^{-1}$. The dividing luminosity is chosen such that the
two resulting subsamples contain similar numbers of
sources (37 and 40), which maximizes the statistical quality of the
analysis. We obtain drastically different absorption 
distributions for the low-luminosity and high-luminosity subsamples,
as shown in Fig.~\ref{nhdist_lum}. While two thirds of AGNs with
$\lxint<10^{43.5}$~erg~s$^{-1}$ are absorbed
($10^{22}$~cm$^{-2}<\nh<10^{24}$~cm$^{-2}$), the corresponding
fraction among those with $\lxint>10^{43.5}$~erg~s$^{-1}$ is less than 20\% 
(2$\sigma$ upper limit). We emphasize once again that we do not
consider Compton thick sources. We also note that since the difference between 
observed and intrinsic luminosity is fairly small,
$\log(\lxint/\lx)=0.1$ (0.5) for $\nh=10^{23}$~($10^{24}$)~cm$^{-2}$,
the above result will essentially not change if the division of sources at
$10^{43.5}$~erg~s$^{-1}$ is done in terms of the observed
luminosity. 

We point out that the tight upper limit obtained above on the fraction of
absorbed high-luminosity AGNs is linked to the fact that there is
only one source (Seyfert 2 galaxy NGC~7582) with
$\lxint>10^{43.5}$~erg~s$^{-1}$ and $\nh>10^{22}$~cm$^{-2}$ in our
sample. We recall however that we simply assumed $\nh=0$ for
those 11 sources (all are optically type 1s) for which only an upper
limit on $\nh$ exceeding $10^{22}$~cm$^{-2}$ is available from
observations. Seven of these sources have
$\lxint>10^{43.5}$~erg~s$^{-1}$. There is a non-negligible probability
that 1 or 2 of these sources are absorbed with
$\nh>10^{22}$~cm$^{-2}$. However, even then the fraction of
absorbed AGNs with $\lxint>10^{43.5}$~erg~s$^{-1}$ will remain small: $10\pm
8$\% and $14\pm 7$\%, respectively.

\subsection{Comparison with other studies}
\label{nh_dist_comp}

Until recently, our knowledge of the distribution of AGNs in intrinsic
absorption was restricted to the local Universe and originated from optically
selected samples. In particular, it was known that Seyfert 2 galaxies
outnumber their Seyfert 1 counterparts by a factor of 4
\citep{mairie95}. Also, \citet{risetal99} estimated the column density
distribution of Seyfert 2s using a sample limited in the
intensity of the [O iii] 5007 \AA\, narrow emission line taken as an
indicator (after some correction) of the intrinsic AGN flux. These
authors concluded that Compton thick ($\nh>10^{24}$~cm$^{-2}$) and Compton thin
($\nh<10^{24}$~cm$^{-2}$) Seyfert 2s are approximately equally  
abundant. Since only a small fraction of Seyfert 1s exhibit
absorbed X-ray spectra, the above papers imply a 2:1 ratio of Compton thin
absorbed ($10^{22}$~cm$^{-2}<\nh<10^{24}$~cm$^{-2}$) to
unabsorbed ($\nh<10^{22}$~cm$^{-2}$) AGNs. This is in good agreement with our 
result for low-luminosity ($\lx<10^{43.5}$~erg~s$^{-1}$)
AGNs. However, our result for the ratio of strongly absorbed 
($10^{23}$~cm$^{-2}<\nh<10^{24}$~cm$^{-2}$) to moderately absorbed
($10^{22}$~cm$^{-2}<\nh<10^{23}$~cm$^{-2}$) low-luminosity AGNs of
$4.1^{+2.9}_{-1.6}$ differs from that of Risaliti et al. of $1.7\pm
0.7$ (the errors are 1$\sigma$), but not significantly. 
Since the Risaliti et al. sample consists almost entirely of
AGNs of low luminosity, the completely different $\nh$ distribution that 
we infer for AGNs with $\lx>10^{43.5}$~erg~s$^{-1}$ could not be
assessed on its basis. 

More recently, medium-sensitivity and deep X-ray surveys have begun to
provide statistical information on the $\nh$ distribution of AGNs. In
particular, utilizing a sample constructed from several surveys
performed in the standard 2--10~keV band with HEAO-1, ASCA and Chandra,
\citet{uedetal03} came to the same conclusion that we reach here that the
fraction of absorbed (Compton thin) AGNs decreases with
luminosity. Furthermore, the fraction of absorbed  sources among
low-luminosity ($\lx\lesssim 10^{43.5}$~erg~s$^{-1}$) AGNs found by
these authors ($\approx 60$\%) is in good agreement with our estimate
($68\pm 8$\%). On the other hand, their estimated value of $\sim
30$--40\%  for this fraction among higher-luminosity AGNs is only
marginally consistent with our 2$\sigma$ upper limit of 20\% (possibly
30\% if 1 or 2 of the 7 type 1 AGNs in our sample with upper 
limits on the column density have $\nh>10^{22}$~cm$^{-2}$, see
\S\ref{nh_dist_lum}). However, the results of Ueda et al. quoted
above are obtained for a heterogeneous sample combining local and very
distant AGNs ($z=0.01$--3), while our estimates are made for the local AGN
population. The apparent discrepancy may therefore hint at a 
substantial cosmological evolution of the intrinsic absorption distribution of
powerful, quasar-like AGNs. 

In another work, \citet{steetal03} investigated the fraction of
optically identified broad-line AGNs among X-ray sources detected at
2--8~keV with Chandra and ASCA. For the $z=0.1$--1 
population, this fraction was found to increase with luminosity, from
less than 50\% at $\lx\lesssim 10^{43}$~erg~s$^{-1}$ to more than
$85$\% at $\lx\gtrsim 10^{44}$~erg~s$^{-1}$, with most of the
uncertainty resulting from the large number of unidentified sources,
especially at low luminosity. If we associate broad-line sources with
X-ray unabsorbed AGNs, the Steffen et al. results for the $z=0.1$--1
population appear to be consistent with ours for the local one. 

\section{X-ray AGN luminosity function}
\label{lum_func}

We now address the X-ray luminosity function of nearby AGNs. We define
this function $\phi(\lx)$  as the number density of AGNs per
$\log\lx$, where $\lx$ is the observed luminosity in the 3--20~keV range.

We first estimate $\phi(\lx)$  in binned form using the conventional 
$1/\vm$ method \citep{schmidt68}. Here, as in the previous
section, $\vm (\lx,\nh)$ is the space volume over which a given AGN
with its observed luminosity $\lx$ and estimated absorption column $\nh$
(again assuming an intrinsic power-law spectrum with $\Gamma=1.8$)
could be detected.

As a next step, we approximate the data by a smoothly connected two
power-law model
\beq
\phi(\lx)=\frac{A}{(\lx/\lb)^{\gamma_1}+(\lx/\lb)^{\gamma_2}}.
\label{lum_model}
\eeq
For this purpose, the maximum likelihood estimator
\beq
L=-2\sum_{j}\ln\frac{\phi(L_{{\rm 3-20},j})V_0(L_{{\rm 3-20},j})}
{\int\phi(\lx)V_0(\lx)\,d\log\lx}
\label{like_lum}
\eeq
is used, where $j$ goes over all sampled AGNs.

The sampled volume $V_0(\lx)$ introduced above derives from the 
previously defined $\vm(\lx,\nh)$ as follows:
\beqa
V_0(\lx) &=& \int f(\nh) \vm(\lx,\nh)\,d\nh
\nonumber\\
&=& \vm(\lx,\nh=0)
\nonumber\\
&&\times
\left\{
\begin{array}{ll}
0.78, & \lx<10^{43.5}\,{\rm erg}\,{\rm s}^{-1}\\
1, & \lx>10^{43.5}\,{\rm erg}\,{\rm s}^{-1}
\end{array}
\right.
\label{vol_corr}
\eeqa
According to this formulation, the $\nh$ values estimated for the individual
AGNs in our sample are not taken into account, but allowance is
made for the average absorption distribution of AGNs $f(\nh)$ as a function of
luminosity derived in \S\ref{nh_dist}. It means that we evaluate the
probability of observing an AGN with a given luminosity $\lx$,
regardless of its intrinsic absorption. We point out that the binned
luminosity function obtained using the $1/\vm$ method inherently takes
into account individual column densities.

Minimizing $L$ yields the best-fit values of the break luminosity $\lb$
as well as of the slopes $\gamma_1$ and $\gamma_2$. The fitting
procedure does not however allow us to determine the normalizing
constant $A$ of the best-fit model. We thus calculate $A$ from the condition
that the number of AGNs predicted by the model is equal to the actual
number of AGNs in our sample.

\begin{figure}
\centering
\includegraphics[width=\columnwidth]{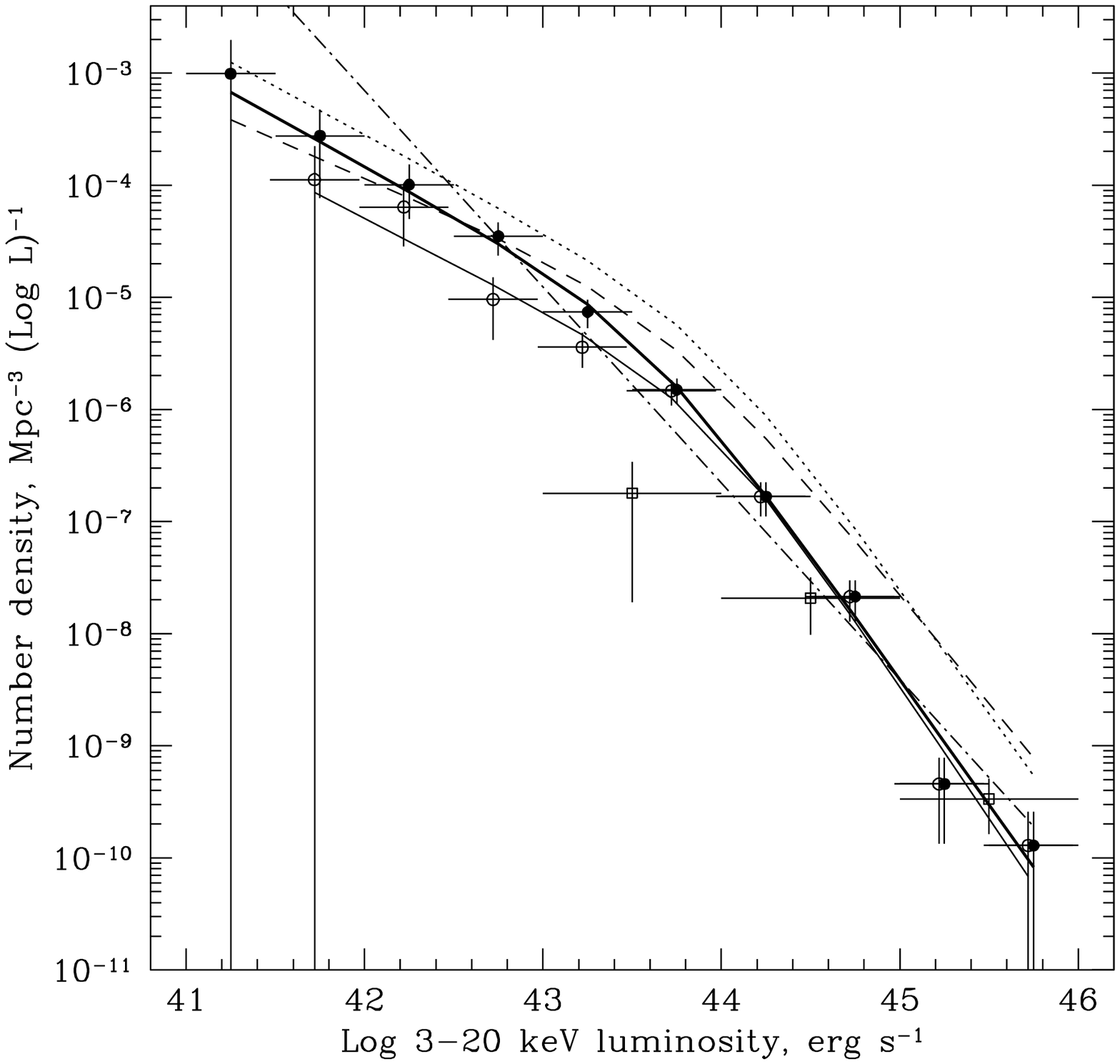}
\caption{Local 3--20~keV luminosity functions obtained here for (i)
Compton-thin ($\nh<10^{24}$~cm$^{-2}$) non-blazar AGNs:  in
binned form (solid circles with 1$\sigma$ error bars) and an analytic
approximation (thick solid curve) given by Eq.~(\ref{lum_model}) with
parameter values quoted in Table~\ref{lumfunc_table}, (ii) similarly 
for unabsorbed ($\nh<10^{22}$~cm$^{-2}$) non-blazar AGNs (open
circles and thin  solid curve), and (iii) for blazars (open
squares within wider bins, no analytic fit is presented). The second set
of data points is  slighlty shifted along the luminosity axis for
better visibility. For comparison are presented previous estimates of
the local 2--10~keV AGN luminosity function, recomputed here to the
3--20~keV band and to our adopted value $H_0=75$~km~s$^{-1}$~Mpc$^{-1}$: by 
\citet{picetal82} (dash-dotted curve), by \citet{uedetal03} (for
Compton-thin sources, dotted curve) and by \citet{lafetal02} (for
unabsorbed sources, dashed curve). 
\label{lumfunc}
} 
\end{figure}

We perform calculations for the luminosity range
$10^{41}$~erg~s$^{-1}<\lx< 10^{46}$~erg~s$^{-1}$, which includes all of
our emission-line AGNs except NGC~4945. The estimated
luminosity of NGC~4945 is $8\times 10^{40}$~erg~s$^{-1}$ and 
as was noted in \S\ref{sample}, this is the only Compton thick AGN in
our sample while the subject of our study is Compton thin AGNs. We
present the derived binned luminosity function and best-fit model in
Fig.~\ref{lumfunc}. The parameters of the model together with their
estimated 1$\sigma$ statistical uncertainties are summarized in
Table~\ref{lumfunc_table}. The normalization $A$ is given without an
error because this parameter is strongly correlated with the
others. The analytic fit is apparently in good agreement with the
binned $\phi(\lx)$; the good quality of the fit is confirmed by the
Kolmogorov-Smirnov test.

\begin{table*}
\caption{3--20 keV AGN luminosity function parameters}
\smallskip

\begin{tabular}{lllllllllr}
\hline
\hline
\multicolumn{1}{c}{Sample} &
\multicolumn{1}{c}{$\log L_{3-20}$} &
\multicolumn{1}{c}{Size} &
\multicolumn{1}{c}{$\log L_\ast$$^{\rm a}$} &
\multicolumn{1}{c}{$\gamma_1$} &
\multicolumn{1}{c}{$\gamma_2$} & 
\multicolumn{1}{c}{$A$$^{\rm b}$} & 
\multicolumn{1}{c}{$W_{3-20}(>41)$} & 
\multicolumn{1}{c}{$N_{3-20}(>41)$} &
\multicolumn{1}{c}{$P_{\rm KS}$$^{\rm c}$} \\

\multicolumn{1}{c}{} &
\multicolumn{1}{c}{erg s$^{-1}$} &
\multicolumn{1}{c}{} &
\multicolumn{1}{c}{erg s$^{-1}$} &
\multicolumn{1}{c}{} &
\multicolumn{1}{c}{} &
\multicolumn{1}{c}{Mpc$^{-3}$} &
\multicolumn{1}{c}{10$^{38}$ erg s$^{-1}$ Mpc$^{-3}$} &
\multicolumn{1}{c}{10$^{-4}$ Mpc$^{-3}$} &
\multicolumn{1}{c}{} \\
\hline

Total & 41--46 & 76 & $43.58_{-0.30}^{+0.32}$ & $0.88_{-0.20}^{+0.18}$ & 
$2.24_{-0.18}^{+0.22}$ & $6.0 (8.6)\times 10^{-6}$ & 
$4.3_{-0.6}^{+0.7}$ ($6.1_{-0.9}^{+1.0}$) & $5_{-2}^{+4}$ ($7_{-3}^{+6}$) & 
$>0.9$ \\

Total & 41--45 & 73 & $43.52_{-0.40}^{+0.40}$ & $0.86_{-0.24}^{+0.20}$ & 
$2.16_{-0.26}^{+0.36}$ & 
$7.3\times 10^{-6}$ & $4.3_{-0.6}^{+0.7}$ & $5_{-2}^{+4}$ & $>0.9$ \\

Unabsorbed & 41--46 & 59 & $43.84_{-0.26}^{+0.26}$ & $0.74_{-0.20}^{+0.20}$ & 
$2.32_{-0.22}^{+0.26}$ & 
$1.9\times 10^{-6}$ & $1.9_{-0.3}^{+0.4}$ & $1.4_{-0.6}^{+1.5}$ & $>0.9$\\

North & 41--45 & 33 & $43.78_{-0.40}^{+0.44}$ & $0.96_{-0.24}^{+0.20}$ & 
$2.26_{-0.26}^{+0.34}$ & $2.5(2.9)\times 10^{-6}$ & 
$3.8_{-1.0}^{+1.1}$ ($4.4_{-1.2}^{+1.3}$) & $5_{-2}^{+5}$ ($6_{-2}^{+6}$) & 
0.88\\

South & 41--46 & 43 & $43.24_{-0.42}^{+0.50}$ & $0.66_{-0.48}^{+0.34}$ & 
$2.14_{-0.22}^{+0.32}$ & $2.2(3.7)\times 10^{-5}$ & 
$4.5_{-0.8}^{+1.0}$ ($7.5_{-1.3}^{+1.7}$) & $4_{-2}^{+4}$ ($7_{-3}^{+7}$) & 
$>0.9$\\

\end{tabular}

$^{\rm a}$ -- All presented uncertainties are 1$\sigma$.

$^{\rm b}$ -- In parentheses are given values corrected for maximum possible
sample incompleteness.

$^{\rm c}$ -- Kolmogorov--Smirnov probability.

\label{lumfunc_table}
\end{table*}

The RXTE slew survey is effectively limited by a redshift $z=0.1$ for
AGNs with $\lx<10^{44.5}$~erg~s$^{-1}$. However, our sample of 
emission-line AGNs includes 6 sources located at $z=0.1-0.3$, all of
which have $\lx>10^{44.7}$~erg~s$^{-1}$. Nevertheless, we are
confident that the obtained luminosity function is characteristic
of the local Universe at $z<0.1$. To prove this, we repeat the fitting in the
narrower luminosity range
$10^{41}$~erg~s$^{-1}<\lx<10^{45}$~erg~s$^{-1}$, so that the
fraction of AGNs with $z>0.1$ reduces to 3 out of 7 in the interval
$10^{44.5}$~erg~s$^{-1}<\lx<10^{45}$~erg~s$^{-1}$. As can be seen in 
Table~\ref{lumfunc}, the best-fit model remains essentially unchanged. 

We also present in Fig.~\ref{lumfunc} and
Table~\ref{lumfunc_table} the luminosity function 
obtained for unabsorbed ($\nh<10^{22}$~cm$^{-2}$) emission-line
AGNs. In calculating the best-fit model in this case, the volume
correction given by equation~(\ref{vol_corr}) was not 
made. As could have been expected from the behavior of the $\nh$
distribution with luminosity (\S\ref{nh_dist}), the contribution of
unabsorbed sources to the number density of Compton thin AGNs is
smaller than 50\% at $\lx<\lb\approx 10^{43.5}$~erg~s$^{-1}$, but
becomes dominant at $\lx>\lb$. It is interesting that the
dramatic change in the intrinsic absorption distribution with
luminosity that we discussed in \S\ref{nh_dist} seems to take
place somewhere near the break of the AGN luminosity function, although our  
sample is not large enough to follow in detail the absorption distribution as a
function of luminosity. 
  
We have further estimated the X-ray luminosity function of blazars,
disregarding the fact that only the 9 lower-luminosity
($\lx<10^{45}$~erg~s$^{-1}$) blazars in our sample  belong to the 
local population ($z\lesssim 0.1$) of AGNs. Interestingly, as
demonstrated by Fig.~\ref{lumfunc}, the number density of blazars
becomes comparable to that of normal (emission-line) AGNs at observed
luminosities above $\sim 10^{44}$~erg~s$^{-1}$. It is worth noting 
that our blazars, selected in the 3--20~keV band, have spectra (see Paper~1)
characterized by a broad distribution of slopes ($\Gamma\sim 1-3$)
centered near the canonical $\Gamma=1.8$ value for unabsorbed
emission-line AGNs.

\subsection{Volume emissivity}
\label{emis}

\begin{figure}
\centering
\includegraphics[width=\columnwidth]{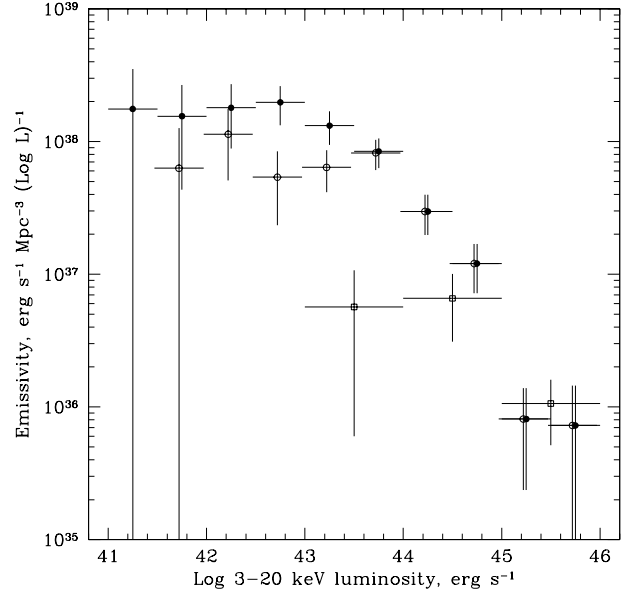}
\caption{Contribution of AGNs with various luminosities to the local
3--20~keV volume emissivity, estimated from the binned luminosity functions
presented in Fig.~\ref{lumfunc}: of Compton-thin non-blazar AGNs (filled
circles), of unabsorbed non-blazar AGNs (open circles), and of blazars (open
squares). 
\label{lumfunc_en}
} 
\end{figure}

Multiplying the luminosity functions shown in Fig.~\ref{lumfunc} by
the luminosity yields a new plot (Fig.~\ref{lumfunc_en}) that allows
one to compare the contributions of AGNs with various
luminosities to the local X-ray volume emissivity. We can see that
AGNs, mostly absorbed ones, with luminosities in the range
$10^{41}$--$10^{43.5}$~erg~s$^{-1}$ release similar amounts of energy in
X-rays per $\log \lx$. We can integrate once more over luminosity,
\beq
\wx(>41)=\int_{41}^\infty \phi(\lx)\lx\, d\log\lx,
\eeq
to estimate the cumulative emissivity of emission-line AGNs with
$\lx>10^{41}$~erg~s$^{-1}$. The resulting value is presented in
Table~\ref{lumfunc_table} together with a 1$\sigma$ statistical
uncertainty determined by exploring the likelihood distribution over
the parameter space. It can be seen that $\wx(>41)$ is well
constrained by the data. 
 
For some applications, the total number
density of AGNs with $\lx>10^{41}$~erg~s$^{-1}$, 
\beq
\nx(>41)=\int_{41}^\infty \phi(\lx)\, d\log\lx,
\eeq
also might be important. Our estimate for this quantity is presented in
Table~\ref{lumfunc_table}. We point out that the quoted number density
is dominated by AGNs located near the lower end of the sampled
luminosity range and may increase by a substantial factor
if AGNs with $\lx<10^{41}$~erg~s$^{-1}$ are counted.

\subsection{Additional checks of the results}
\label{check}

The 3--20~keV AGN luminosity function obtained above may be affected
by both the incompleteness of the input sample and the inhomogeneous
distribution of AGNs resulting from the local large scale structure. To
address the first of these issues, we assumed that AGNs meeting 
the XSS detection criterion and not appearing in our sample have the
same distribution in luminosity as the AGNs composing the   
sample. We can then estimate the maximum possible effect of sample
incompleteness by correcting the best-fit model amplitude $A$ as well as the
inferred integral quantitites $\wx(>41)$ and $\nx(>41)$ using the 
completeness coefficient defined in \S\ref{compl}. The corrected
amplitude is given in Table~\ref{lumfunc_table}. The true amplitude of
the luminosity function should lie somewhere between the uncorrected and
corrected values.

\begin{figure}
\centering
\includegraphics[width=\columnwidth]{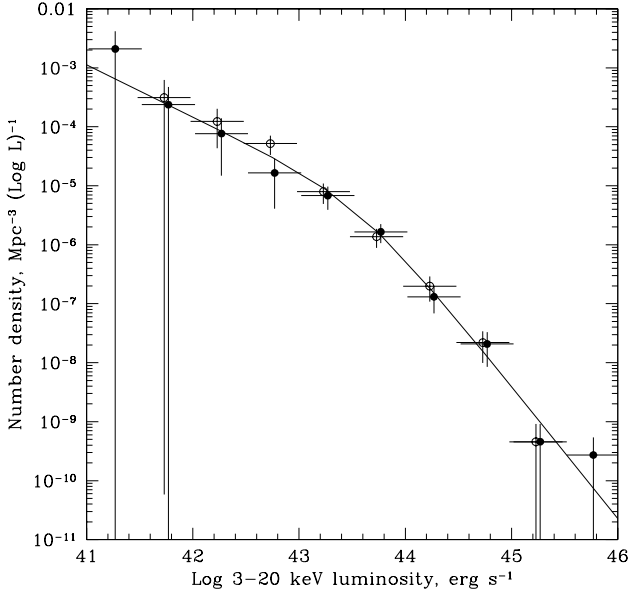}
\caption{3--20 keV luminosity functions of local Compton-thin
non-blazar AGNs inferred from the southern subsample (filled circles)
and northern subsample (open circles), in comparison with the best-fit
model [Eq.~(\ref{lum_model}), Table~\ref{lumfunc_table}] found for the
whole AGN sample (solid line).  
\label{lumfunc_ns}
} 
\end{figure}

In order to estimate the possible effect of the large scale structure,
we have computed luminosity functions for our northern and southern
subsamples of AGNs. The obtained binned distributions and
corresponding best-fit analytic models are presented in 
Fig.~\ref{lumfunc_ns} and Table~\ref{lumfunc_table}. In addition, the
amplitudes corrected for the incompleteness of both samples are
given. The Kolmogorov--Smirnov test demonstrates that the two AGN
subsamples could well be drawn from the same luminosity
distribution: $P_{\rm KS}=0.35$. Furthermore, the estimated cumulative
emissivity $\wx(>41)$ and number density $\nx(>41)$ are not significantly
different for the northern and southern samples either.

\begin{figure}
\centering
\includegraphics[width=\columnwidth]{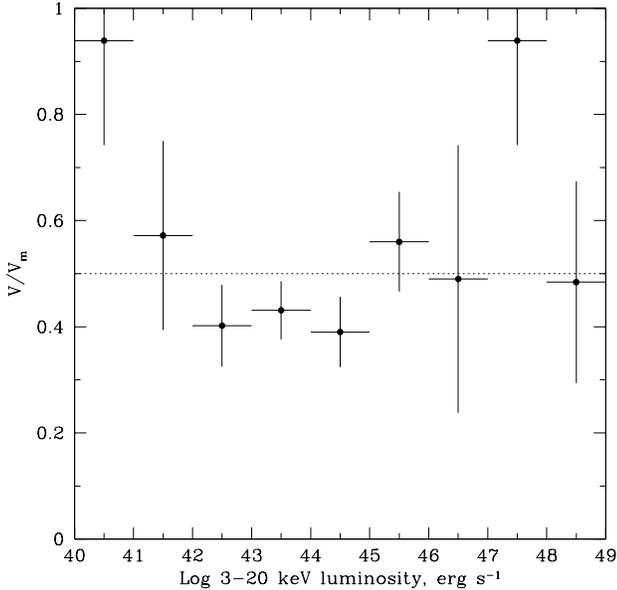}
\caption{$V/\vm$ ratio averaged over luminosity bins for the whole XSS
AGN sample. The error bars represent 1$\sigma$ statistical uncertainties.  
\label{lum_vvmax_bin}
} 
\end{figure}

Finally, we perform a standard $V/\vm$ test \citep{schmidt68} for our
AGN sample. To this end, a $V/\vm$ ratio is calculated for
each AGN as $(C/4\delta C)^{-3/2}$, where $C$ is the measured count
rate in the 3--20~keV band and $\delta C$ is the associated 1$\sigma$
statistical uncertainty. The factor 4 reflects the detection
criterion defining the AGN sample. Then, averaging is performed in
specified luminosity bins. As can be seen from
Fig.~\ref{lum_vvmax_bin}, the observed deviations of the $V/\vm$ ratio
from the value 0.5 expected for a homogeneous distribution of sources
are not statistically significant. In particular, averaging over the
whole sample of emission-line AGNs gives $\langle V/\vm\rangle=0.44\pm
0.04$, which represents an insignificant ($1.5\sigma$) deviation
from 0.5. 

We conclude that the obtained luminosity function is not
significantly affected by the inhomogeneity of matter distribution in the local
Universe. The possible incompleteness of our sample introduces a
systematic uncertainty of $\sim 20$\% on the amplitude of the
luminosity function, characterized by $\wx(>41)$, in addition to the
statistical uncertainty of $\sim 15$\%. 

\subsection{Comparison with other studies}
\label{lum_func_comp}

The 3--20~keV AGN luminosity function obtained in this work surpasses
in quality previously published luminosity functions obtained at photon
energies above 2~keV for the local ($z<0.1$) AGN population. In
Fig.~\ref{lumfunc}, we show for comparison the 2--10~keV luminosity
function derived by \citet{picetal82} from their HEAO-1/A2 sample of AGNs 
(mostly unabsorbed ones). The small luminosity correction
$\log(\lx/\ls)=0.1$ has a negligible effect on the comparison. Since the
Piccinotti sample is relatively small ($\sim 40$\% of ours) and covers
a relatively narrow luminosity range
$10^{42}$~erg~s$^{-1}<\ls<10^{45}$~erg~s$^{-1}$, the data could be
well fitted by a power-law model with an index $1.75$. Within the 
uncertainties, the Piccinotti et al. luminosity function is in very
good agreement with the one derived here. 

In addition, much effort has been invested into studying the
evolution of the 2--10~keV AGN luminosity function with redshift. In
particular, \citet{lafetal02} and \citet{uedetal03} have constructed 
large samples of AGNs for this purpose. The former includes 158
optically type 1 AGNs selected from HEAO-1, BeppoSAX and ASCA surveys,
and the latter consists of 247 AGNs detected by HEAO-1, ASCA and Chandra,
including a substantial number of X-ray absorbed sources. In these
works, both the shape of the luminosity function and its evolution
out to high redshift are fitted simultaneously to the observed
distribution of AGNs on the luminosity-redshift plane. It is 
interesting to compare the present-day luminosity functions predicted
by these studies with the one directly determined here. 

\citet{uedetal03} present their results in terms of the 
intrinsic 2--10~keV luminosity $\lsint$. Although our
luminosity function is defined in terms of the observed 3--20~keV
luminosity $\lx$, it will hardly change after recalculation in terms of
$\lxint$ since $\langle\log(\lxint/\lx)\rangle=0.1$ (0) for the observed
$\nh$ distribution of AGNs with $\lx<10^{43.5}$~erg~s$^{-1}$
($\lx>10^{43.5}$~erg~s$^{-1}$). The similarly small
$\log(\lxint/\lsint)=0.1$ correction counteracts the previous 
one at low luminosity. We can therefore take approximately
$\log(\lx/\lsint)=0.05$ for the entire luminosity 
range. On the other hand, the luminosity function of
\citet{lafetal02} is obtained for unabsorbed AGNs and should thus be
compared with our corresponding result for this case [making a small
correction $\log(\lx/\ls)=0.1$].

The comparison is done in Fig.~\ref{lumfunc}, and it can be seen that
the amplitude of our luminosity functions derived for all AGNs and for
unabsorbed ones is smaller by at least a factor of 2 than predicted
for $z=0$ by \citet{uedetal03} and \citet{lafetal02},
respectively. Part of this apparent discrepancy (less than a factor of 1.4
and this is already reflected in Table~\ref{lumfunc_table}) may result
from the possible incompleteness of our sample. It is difficult to
make a formal statement as to whether the remaining difference can be
accounted for by the statistical uncertainties. Despite the different
amplitudes, the shapes of our luminosity function and those of 
\citet{uedetal03} and \citet{lafetal02} are in satisfactory agreement. 

\subsection{8--20~keV luminosity function}
\label{8_20}

Our preceeding analysis was based on a sample of AGNs selected by
the flux in the 3--20~keV band. At the same time, we know the source fluxes
in the subbands 3--8~keV and 8--20~keV. In addition,
as explained in Paper~1, the sensitivity of the RXTE slew survey
as a function of photon energy (determined by the energy response of
the PCA instrument and by the background) is such that Compton thin
AGNs ($\nh<10^{24}$~cm$^{-2}$) detectable at 8--20~keV are always also
detectable in the broader band 3--20~keV. We can therefore define a
sample of AGNs selected in the 8--20~keV band from our 3--20~keV
sample (Table~\ref{agn_table}) by applying the condition $C_{8-20}/\delta
C_{8-20}>4$, where $C_{8-20}$ and $\delta C_{8-20}$ are the count rate
and 1$\sigma$ uncertainty for the hard subband. 

The hard X-ray selected sample consists of 45 AGNs, including 37
emission-line AGNs (24 unabsorbed and 13 absorbed) and 8
blazars. Among the 35 unidentified sources (Table~\ref{noid_table})
only 7 are detected in the 8--20~keV band, and therefore the new AGN
sample is at least 87\% complete.

\begin{table*}
\caption{8-20 keV AGN luminosity function parameters}
\smallskip

\begin{tabular}{lllllllllr}
\hline
\hline
\multicolumn{1}{c}{Sample} &
\multicolumn{1}{c}{$\log L_{3-20}$} &
\multicolumn{1}{c}{Size} &
\multicolumn{1}{c}{$\log L_\ast$$^{\rm a}$} &
\multicolumn{1}{c}{$\gamma_1$} &
\multicolumn{1}{c}{$\gamma_2$} & 
\multicolumn{1}{c}{$A$} & 
\multicolumn{1}{c}{$W_{8-20}(>41)$} & 
\multicolumn{1}{c}{$N_{8-20}(>41)$} &
\multicolumn{1}{c}{$P_{\rm KS}$} \\

\multicolumn{1}{c}{} &
\multicolumn{1}{c}{erg s$^{-1}$} &
\multicolumn{1}{c}{} &
\multicolumn{1}{c}{erg s$^{-1}$} &
\multicolumn{1}{c}{} &
\multicolumn{1}{c}{} &
\multicolumn{1}{c}{Mpc$^{-3}$} &
\multicolumn{1}{c}{10$^{38}$ erg s$^{-1}$ Mpc$^{-3}$} &
\multicolumn{1}{c}{10$^{-4}$ Mpc$^{-3}$} &
\multicolumn{1}{c}{} \\
\hline

Total & 40.5--44.5 & 37 & $42.82_{-0.50}^{+0.52}$ & $0.66_{-0.42}^{+0.32}$ & 
$1.98_{-0.28}^{+0.36}$ & 
$4.6(5.3)\times 10^{-6}$ & $3.6_{-0.6}^{+0.8}$ ($4.3_{-0.7}^{+1.0}$) & 
$10_{-5}^{+10}$ ($12_{-6}^{+12}$) & $>0.9$\\
\hline

\end{tabular}

$^{\rm a}$ -- All presented uncertainties are 1$\sigma$.

\label{lumfunc_8_20_table}
\end{table*}

\begin{figure}
\centering
\includegraphics[width=\columnwidth]{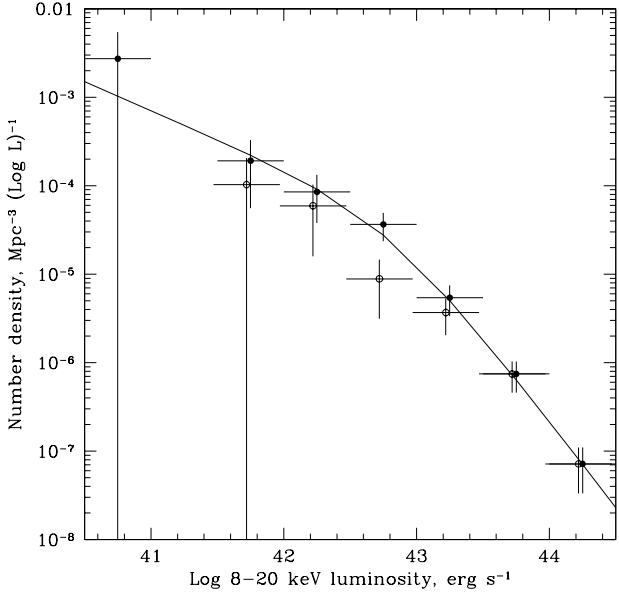}
\caption{Local 8--20~keV luminosity functions obtained here for 
Compton-thin ($\nh<10^{24}$~cm$^{-2}$) non-blazar AGNs:  in
binned form (solid circles with 1$\sigma$ error bars) and an analytic
approximation (thick solid curve) given by Eq.~(\ref{lum_model}) with
parameter values quoted in Table~\ref{lumfunc_table}, and for unabsorbed
($\nh<10^{22}$~cm$^{-2}$) non-blazar AGNs (open circles, no analytic
fit is presented). 
\label{lumfunc_8_20}
} 
\end{figure}

Using the above sample, we have built a 8--20~keV AGN luminosity
function in the sampled range
$10^{40.5}$~erg~s$^{-1}>\lh<10^{44.5}$~erg~s$^{-1}$ (here $\lh$ is the
observed 8--20~keV luminosity) by the methods described earlier in
this section. The results are presented in Fig.~\ref{lumfunc_8_20} and
Table~\ref{lumfunc_8_20_table}. In  calculating the best-fit model we
made a small volume correction to take into account the difference in
the $\nh$ distribution of low-luminosity 
and high-luminosity AGNs:
\beqa
V_0(\lh) &=&\vm(\lh,\nh=0)
\nonumber\\
&&\times\left\{
\begin{array}{ll}
0.84, & \lh<10^{43.2}\,{\rm erg}\,{\rm s}^{-1}\\
1, & \lh>10^{43.2}\,{\rm erg}\,{\rm s}^{-1}
\end{array}
\right.
\label{vol_corr_8_20}
\eeqa
This expression differs from equation~(\ref{vol_corr}) because the
dependence of the survey volume on $\nh$ is different for the
3--20~keV and 8--20~keV bands (see Fig.~\ref{vol_nh}), and since
$\log(\lx/\lh)= 0.3$ for $\nh=0$ [for comparison,
$\langle\log(\lx/\lh)\rangle=0.2$ for an ensemble of AGNs with
$\lx<10^{43.5}$~erg~s$^{-1}$ and the derived  $\nh$ distribution].

The small size of our 8--20~keV selected sample compared to the 3--20~keV
selected one leads to relatively large uncertainty in the determination of
the 8--20~keV luminosity function. Given this fact and taking into
account the $\lx/\lh$ correction, the best-fit model obtained for the 8--20~keV
luminosity function (Table~\ref{lumfunc_8_20_table}) is consistent with the
better constrained 3--20~keV luminosity function (Table~\ref{lumfunc_table}).

\section{AGN contribution to the local X-ray output}
\label{energy}

In the previous section, we estimated the total energy released per
unit volume in the 3--20~keV band by local AGNs with
$\lx>10^{41}$~erg~s$^{-1}$ at $\wx(>41)= (5.2\pm 1.2)\times
10^{38}$~erg~s$^{-1}$~Mpc$^{-3}$. The value and error given here take  
into account both the possible sample incompleteness and statistical
uncertainty. Given the luminosity dependence of the absorption column
density distribution discussed in \S\ref{nh_dist}, we can convert the 
above estimate to the standard 2--10~keV band: $\ws(>40.7)=
(2.9\pm 0.7)\times 10^{38}$~erg~s$^{-1}$~Mpc$^{-3}$. Most of this
emission is produced by Seyfert galaxies with X-ray luminosities below
$\sim 10^{44}$~erg~s$^{-1}$.

It is interesting to compare the above value with the total X-ray
volume emissivity in the 2--10~keV band, which has been estimated as
$\rho_{\rm 2-10}=(6.5\pm 1.9)\times 10^{38}$~erg~s$^{-1}$~Mpc$^{-3}$ (for
$H_0=75$~km~s$^{-1}$~Mpc$^{-1}$) from the cross-correlation of IRAS
galaxies with the HEAO-1 all-sky X-ray map    
\citep{miyetal94}. We may conclude from this comparison that sources other
than classical Seyfert galaxies may provide similar contribution to
the local X-ray emissivity. The obvious candidates are low-luminosity
($\ls< 10^{41}$~erg~s$^{-1}$) AGN, starburst and non-active galaxies and
clusters of galaxies. In particular, the contribution of clusters of
galaxies  to the local 2--10~keV emissivity is $\sim 0.5\times
10^{38}$~erg~s$^{-1}$~Mpc$^{-3}$, as can be estimated from
the measured 0.1--2.4~keV luminosity function \citep{boeetal02} and the
luminosity--temperature relation \citep{markevitch98}.

We may further compare the local AGN volume emissivity with relevant
estimates for a more distant Universe. The results of 
\citet{cowetal03} obtained with the Chandra observatory indicate that
the cumulative emission from AGNs with $\ls>10^{42}$~erg~s$^{-1}$ (the
actual energy range used was 2--8~keV) has decreased from a few $\times 
10^{39}$~erg~s$^{-1}$~Mpc$^{-3}$ at $z=1$--2 to less than
$10^{39}$~erg~s$^{-1}$~Mpc$^{-3}$ at $z\sim 0.5$. Our result thus suggests
that a further decrease of the total energy production by AGNs has occured
by the present epoch.

\section{Conclusions}
\label{conc}

\begin{enumerate}

\item
A well defined sample of 95 AGNs located at $|b|>10^\circ$, detected 
in the 3--20~keV band by the RXTE slew survey is presented. Most of
the sources belong to the local population ($z<0.1$). Accurate
estimates of the intrinsic absorption column are presented for
practically all of the sources.

\item
The reconstructed $\nh$ distribution of AGNs is drastically different
for low-luminosity ($\lx\lesssim 10^{43.5}$~erg~s$^{-1}$) and
high-luminosity ($\lx\gtrsim 10^{43.5}$~erg~s$^{-1}$) objects. Among
the former, two thirds are X-ray absorbed
($10^{22}$~cm$^{-2}<\nh<10^{24}$~cm$^{-2}$), whereas the corresponding
fraction is less than 20\% among the latter. These statistics do not
take into account the population of Compton thick AGNs
($\nh>10^{24}$~cm$^{-2}$), to which our survey is not sufficiently
sensitive. 

\item
The 3--20~keV AGN luminosity function is derived, which is the
best-to-date in quality at photon energies above 2~keV for the
local population. The luminosity function starts to flatten
at $\lx\approx  10^{43.5}$~erg~s$^{-1}$ toward lower luminosities,
approximately where obscured objects start to dominate the AGN
population. A physical explanation for this behavior is required. 

\item
Comparison of the cumulative X-ray output of AGNs with
$\lx>10^{41}$~erg~s$^{-1}$ with the previously estimated total X-ray
volume emissivity in the local Universe demonstrates that
low-luminosity ($\lx<10^{41}$~erg~s$^{-1}$) AGNs, non-active galaxies
and clusters of galaxies together may be emitting a similar amount of
X-rays to Seyfert galaxies. 

\item
A sample of 35 unidentified sources -- AGN candidates, detected during the
RXTE slew survey is presented. 12 of these have a likely ROSAT soft
X-ray counterpart with a better than 1~arcmin localization, so their
identification in the optical and other bands should not be difficult. The
positions of the other sources are currently known with a 1~deg
accuracy, and could be improved by scanning observations with X-ray
telescopes.  

\end{enumerate}

\noindent {\sl Acknowledgments} We thank the anonymous referee for careful
reading and constructive comments.


\end{document}